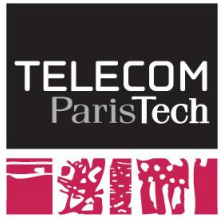

# Data Protection: Combining Fragmentation, Encryption, and Dispersion,
## Final report

Gérard Memmi, Katarzyna Kapusta, Patrick Lambein, and Han Qiu

Télécom ParisTech, CNRS LTCI, Université Paris-Saclay
{gerard.memmi, katarzyna.kapusta, patrick.lambein, han.qiu}@telecom-paristech.fr

# Contenu

## Abstract:

Hardening data protection using multiple methods rather than 'just' encryption is of paramount importance when considering continuous and powerful attacks to observe, steal, or even destroy private and confidential information.

Our purpose is to look at cost effective data protection by way of combining fragmentation, encryption, and then dispersion over several physical machines. This involves deriving general schemes to protect data everywhere throughout a network of machines where they are being processed, transmitted, and stored during their entire life cycle. This is being enabled by a number of parallel and distributed architectures using various set of cores or machines ranging from General Purpose GPUs to multiple clouds.

In this report, we first present a general and conceptual description of what should be a fragmentation, encryption, and dispersion system (FEDS) including a number of high level requirements such systems ought to meet. Then, we focus on two kind of fragmentation. First, a selective separation of information in two fragments a public one and a private one. We describe a family of processes and address not only the question of performance but also the questions of memory occupation, integrity or quality of the restitution of the information, and of course we conclude with an analysis of the level of security provided by our algorithms. Then, we analyze works first on general dispersion systems in a bitwise manner i.e. without data structure consideration; second on fragmentation of information considering data defined along an object oriented data structure or along a record structure to be stored in a relational database. We present an extension of Shamir's algorithm with similar security analysis and conclude that combining data fragmentation, encryption, and dispersion constitutes a potent process to protect data.

## Key Words:

Data protection, information protection, fragmentation, defragmentation, data dispersion, encryption, selective encryption, AES, DCT, DWT, FHE, Shamir algorithm, security analysis, image encryption, General Purpose GPU computation, multilevel security, Cloud Computing, database, privacy

### Résumé (*in French*):

*Nous présentons une description conceptuelle générale d'un système de fragmentation et de défragmentation des données afin de les mieux protéger. Nous identifions de nombreux besoins et propriétés de haut niveau associées à des traits d'architecture qu'un tel système doit adresser et posséder avec un souci de performance et de cout optimisés. Il nous parait aujourd'hui important de combiner tout à la fois fragmentation, chiffrement et dispersion dans une architecture qui se doit de rester efficace avec un cout bien maitrisé.*

*Nous avançons sur plusieurs fronts. D'une part, nous expérimentons avec la protection sélective de données multimédia ou textuelle en développant de nouveaux algorithmes permettant de séparer automatiquement l'information qui sera traitée de façon confidentielle de la donnée qui pourra être traitée de façon publique, par exemple stockée sur un cloud. Nous avons pu mettre au point une famille d'algorithmes de protection sélective des données (images bitmap ou texte) avec des performances supérieures à une encryption complète de l'image sur CPU ou GPGPU. Pour ce faire, nous avons parallélisé le traitement des algorithmes et utilisé des architectures utilisant à la fois le processeur central (CPU) et les processeurs graphiques (GPGPU). Cette famille d'algorithmes propose à l'utilisateur des niveaux de protection différents (fort ou léger) pour des usages de protection adaptés. Nous avons généralisé ces méthodes vers d'autres natures ou formats de données (en particulier, texte et vidéo) en choisissant une transformation intègre (ondelettes) pour séparer données privée et donnée laissées publiques.*

*Pour traiter des données de façon non sélective, nous avons pu développer un état de l'art en deux parties. Le premier comporte six systèmes (trois commerciaux et trois académiques) avec diverses propriétés. Ces systèmes sont caractérisés par le fait qu'ils ne font aucune hypothèse sur la structure de la donnée qu'ils ont à protéger. Leur analyse donne une grille multicritère de lecture et d'évaluation de tout système ayant un modèle similaire. La seconde partie de l'état de l'art présente des systèmes plus anciens et plus classiques pour lesquels l'usager décrit les différents niveaux de confidentialité d'une donnée dont le système connait également la structure (soit orientée objet, soit base de données). Nous avons également travaillé à l'extension de l'algorithme de Shamir pour prendre en compte des données de grande taille, nous avons parallélisé son code pour obtenir des performances acceptables.*

### Mots clés:

Protection des données, protection de l'information, fragmentation, défragmentation, dispersion des données, chiffrement, chiffrement sélectif, AES, DCT, DWT, FHE, algorithme de Shamir, chiffrement d'images, analyse de sécurité, sécurité multiniveau, General Purpose GPU, Cloud Computing, base de données, privacy

# 1. Introduction

With over three decades long, outsourcing information storage and processing, cloud-based services for data storage have gained in popularity and today can be considered as mainstream. They attract organizations or enterprises as well as individual users who do not want or cannot cope with the cost of a private cloud. Beside the economic factor, both groups of customers subordinate their choice of an adequate cloud provider to other factors, particularly resilience, security, and privacy.

Hardening data protection using multiple methods rather than 'just' encryption is becoming of paramount importance when considering continuous and powerful attacks to spy, alter, or even destroy private and confidential information. Even if encryption is a great technology rapidly progressing, encryption is 'just' not enough to progress with this unsolvable question not mentioning its high computational complexity. In [Adrian…15], it is shown how to compromise https sites with 512-bit group; the authors even suggested that 1024- bit encryption could be cryptanalyzed with enough computational power. Cryptographs never like the idea that a cypher can be broken and information can be read given sufficient computational resources, this is nevertheless one of the central design tenets of a number of projects like the Potshards system [Storer…09]. Moreover, there remains the difficult question of the management of the encryption key that over time, can be known by too many people, and stolen or lost.

Our purpose and ultimate ambition is to look at data protection and privacy from end to end by way of combining fragmentation, encryption, and then dispersion. This means to derive general schemes and architecture to protect data during their entire life cycle everywhere they go throughout a network of machines where they are being processed, transmitted, and stored. Moreover, it is to offer users choices among various well understood cost effective levels of privacy and security which would come with predictable levels of performance in terms of memory occupation, energy consumption, and processing time. However, we have to reduce this long term ambition in order to stay within the means and workload defined in the CAP project. For this project our focus will be set on protecting data during their storage. For that purpose, we will assume that our end user will have a resource limited personal environment and will look at a third party storage provider with a cost effectiveness additional constraint.

In the recent past, major enterprises (Sony[1] or Yahoo[2] for instance) have been seriously attacked. Even political organizations are not escaping to hackers and other intelligence organizations. On this basis, it is only natural to assume that at least one 'uninvited' unidentified program is hosted on at least one machine and has acquired the rights to be able to observe, share with other 'uninvited' components, steal, or even destroy data that are processed, transmitted through, or stored on this very machine. This machine is in general part of a distributed system and for instance, can be part of a datacenter or a cloud. We will assume though, that machines where data are entered or presented in clear to an authorized end-user are under his responsibility and can be trusted. In other words, it does not seem reasonable to make information accessible in clear to an authorized end-user and simultaneously on the same machine, not available to an uninvited party or component.

---

[1] https://en.wikipedia.org/wiki/Sony_Pictures_hack
[2] https://www.nytimes.com/2016/12/14/technology/yahoo-hack.html?_r=1

In this report, we first present a conceptual description of what is a general fragmentation and defragmentation system including a number of high level requirements. Then, we focus on fragmentation for protecting data during their storage and plan to take a deeper look at other data life cycle phases (transmission or processing) in another project. We first consider a bitmap image describing results published in [Qiu…14] and enhanced in [Qiu…15]: we discuss previous works on selective encryption methods and their performance issue. We present the designs of our method in particular, introducing a new design which provides a strong level of protection. We discuss their parallel implementations using GPGPU (General Purpose Graphics Processing Unit). Moreover, in order to obtain good performance, the implementation of data fragmentation is performed partly on a CPU, partly on a GPU. We observe as expected that computation is considerably sped up particularly when matrix computation is involved (in line with [Modieginyane…13]). We are analyzing in details performance, security, and integrity issues, and conclude by describing how selective encryption can be used to safely store public fragments in public storage systems.

Then, we analyze works first, on general data dispersion systems in a bitwise manner, comparing academic and commercial solutions. Second, we gather publications on fragmentation in a structurewise manner considering data stored either in an object oriented system or in a structured database, following a survey started in [Kapusta…15].

Beside some standard images that be found on the web, small textual or video samples, we have been experimenting with La Poste data available in the ITEA2 CAP project to explore and experiment with scalability of our methods. More details about our working environment in terms of hardware, software, and data can be found in the Annex, section 12.

### 1.1. Dissemination

Early results of this project were about image protection and were published in [Qiu…14] and expended in [Qiu…15]. Then a state of the art on general fragmentation methods was published [Kapusta…15], and [Memmi…15] and then presented at the FOSAD Summer School (http://www.sti.uniurb.it/events/fosad16/kapusta.pdf). Additionally, two posters were accepted at the RAID'16 Symposium ([Kapusta…16b] and [Qiu…16]) and another one to ACM CCS'16 Conference [Kapusta…16a] expended into a full paper and submitted for 2017 [Kapusta…17]. This report abundantly draws from these various publications. Several other publications are being prepared. At last, several student projects were proposed directly derived from this research and aiming at increasing the number of experiments using the code developed through the project.

## 2. Motivations

### 2.1. Data protection, integrity, and privacy

Data must be protected during their entire life cycle (or at least during the period of time its owner cares about: data is aging and its pertinence can vary over time), not only when they are stored but also when they are transmitted from one location to another during which they can be intercepted and altered; later, when they are archived. Today, datacenters are continuously being attacked, even if strong defense mechanisms are continuously put in place and updated, the risk stays acute and data protection to achieve security and privacy is a must for any user of public or hybrid storage system.

## 2.2. Cost effectiveness

With the increase of sources of data production at an unbelievable pace and the multiplication of possibilities to store them particularly in cost effective and public data storage areas, it becomes quite clear that opportunities to observe (by someone 'honest but curious', in particular like anyone from the hardware provider, the cloud provider, to the Saas provider) or worse to alter, or even destroy valuable data are accordingly multiplied.

## 2.3. Independence

With the diversification of data storage in public and remote areas, it becomes more and more necessary for users to efficiently protect their personal data (texts, images, or videos) by using their own tools *independently* from the ones offered by storage service providers and their affiliates. It seems reasonable to recommend that a user should have a security service provider independent from his other service providers. Otherwise, it would be a little bit like letting the fox guarding the henhouse.

## 2.4. Resilience and availability

Fragmentation is vastly used for resilience purposes. In [Rabin 89], we find one of the first results about fragmenting for both fault-tolerance and data protection. In [Kapusta…17], we address this question by using a Reed Solomon error correcting code [Reed…60] to avoid mere duplication. This solution is also used by many authors.

## 2.5. Performance and scalability

Protecting information can be achieved in many ways: by strictly controlling its access (through authentication techniques or access right policies), hiding it, or transforming it (through techniques like steganography, scrambling, encryption, anonymization, or obfuscation) and making it understandable solely by selected people who own a method, a map, or a key to efficiently retrieve, reverse the protective transformation, and get back the complete original information in a timely manner.

During this process, it is sometimes possible to detect whether data has been accidentally or intentionally modified. These techniques can become extremely costly (in memory and time) especially with large or massive amount of data; hence the idea of adding another weapon to more traditional protection techniques: data fragmentation. Fragmentation means separating with a more or less complex algorithm data into pieces or fragments, considering confidentiality, data nature and size. In order to be cost effective, fragments will be protected differently according to their level of confidentiality or criticality. The nature of data, be a text, an image, or a video will greatly influence the type of analysis, methods and processes that will be performed to be effective. Fragments are then dispersed and stored in several different files. These files should in turn be stored in different physical machines in a more or less sophisticated manner in order to increase the level of protection of the whole information.

It is assumed that massive amount of data have a non-uniform level of criticality or confidentiality (therefore, a non-uniform need for protection). Also, they have not been produced at the same time, they are aging at a non-uniform pace which again relate to the non-uniform level of criticality and a need for a multilevel security system. Some critical data could then be separated and strongly encrypted, some other data less critical would be only fragmented and possibly more rapidly encrypted with a weaker encryption algorithm or even in some use cases, let clear.

Last but not least, by definition, fragmentation enables the parallelization of transforming or encrypting pieces of information which lets us expect strong gain in efficiency compared with a full encryption sequentially executed, addressing scalability requirements. Defragmentation could then have to follow a reverse parallel pattern.

## 2.6. Distributed Storage and Trusted Areas

Whatever is the software solution used for protecting data, it is our belief that a complete solution will have to use hardened hardware (a trusted area of one or several machines) at one critical moment or another during the data life cycle. In particular, places where information is being fragmented or defragmented, encrypted or decrypted are particularly critical since the information is gathered in clear during a period of time. Also, places where information is being created, printed out, or visualized by a human end-user have to be trusted and protected from any uninvited observer. A last reason for considering a trusted area would be to use it as a safe and store ultra-confidential information even as this information is strongly encrypted. This point is widely recognized since a long time and in many publications ([Fray…86] or [Aggarwal…05] for instance) or by many industry experts.

Of course, from a fragmentation point of view, data should not enter and exit this trusted area by a single port in a single iteration which would mean that in front of the trusted area, there would exist a non-trusted location where an observer (a 'man in the middle') can see the entire flow of information. A method has to be designed to gather and transmit data inwards and outwards a trusted area to the rest of the storage system.

Trusted areas have a high cost; one goal for a fragmentation system is to minimize and possibly confine their usage although for the reason said earlier, their usage is and will stay unavoidable if we want to reach an end to end high level of confidentiality.

## 2.7. Use cases

Use cases are important since a specific architecture can comply with a set of use cases but at the same time may very well fail at addressing needs for another group of use cases. Use cases can be defined according to the number of desired authorized participants (one, two, or many), their roles as users or end-users (owner, author (who may not be the owner), read-only user, service provider,…) (aka Alice and Bob), the number and type of attackers (from honest but curious, eavesdropper (aka Eve), to malicious (aka Mallory), insider, man in the middle, coalition of attackers, powerful rogue enterprise,…), the type and location of attacks (at storage, transmission, processing time, …), the size, nature, and format of the data (image, video, text, database, unstructured data,…), the kind of distributed machine environments (one machine to another machine, one personal machine (from a laptop to a mobile device like smartphone or a tablet) to one cloud, a general distributed environment involving several providers,..). We can see by combining these various possibilities that use cases can be very contrasted and their number can be relatively high.

In this project, we consider mainly three use cases two of them with relatively similar solutions. First, an end-user (Alice) wants to save her multimedia data in a public cloud in order to save memory in her private resource-limited environment (be a desktop, a laptop, or even a smartphone), however, for privacy reasons, she does not want putting her entire data in the hands one public storage provider. Second, an enterprise wants to save a massive amount of data balancing confidentiality and cost effectiveness; it is considering using a hybrid cloud solution. Third, we consider a user with a massive amount of data that he wants to disperse over many different machines.

## 3. Fragmenting for protecting

Fragmentation is not a new idea. This concept is used in many domains like in economics or in history. To stay on the safe side, a banker will recommend not 'putting all his eggs in the same basket'. In computer science, the concept of fragmentation can be found in many different applications and usages: by operating system to optimize disk space management, by database management or distributed systems to gain in performance particularly in latency, by routing algorithms in communication to increase reliability and support disaster recovery when combining replication and fragmentation together. A number of widely used technologies can be cited here like Apache Hbase, Google File System, [Raid], etc…

It is also since the invention of writing used to keep a secret. In fact, we could almost consider a cipher as a splitting information algorithm since you need a key as a first fragment of information and an encrypted data a second information fragment in order to recover the original information. We are interested by the usage of any fragmentation technique in combination with encryption and fragment dispersion to protect data of any nature during storage, archiving, or transmission with applications in information privacy.

Reasons for considering or reconsidering fragmentation for protection are manifold: the continuous scaling up of today distributed storage architectures; the massive amount of data produced every day by users or machines, the amazing amount of data and CPU that analytics applications need to gather and process, the need for these data to be transmitted from one server to another. A cloud is supported by a number of interconnected datacenters among which data can be dispersed and stored when latency is not an issue ([Bessani…11]) and a datacenter has a plethora of servers and distributed memory available. We can therefore think about scattering a large number of fragments all over while storing a very limited number of non-contiguous fragments on a single physical machine: the vast number of storage elements, the access to many servers make it worthwhile to revisit the old idea of fragmentation in order to protect data. We started to toy with this kind of motivations back in 2011 [Memmi 11]. A malicious or uninvited observer hidden on a single machine should never see more than a handful of (preferably semantically separated from each other (i.e. the knowledge of a fragment does not clarify the meaning of another one)) fragments and should never be in a position to observe a complete computation. This observer should feel like searching for many needles in many (potentially huge) haystacks. Of course, there is always the additional possibility to confine ultra-critical fragments in a private or trusted storage providing an even higher level of security and privacy.

### 3.1. Fragmentation definition

By information **fragmentation**, we mean the process of constructing a covering (a partition is a special case and in fact, is the case used in this project and the unanimity of the literature) of the information by pieces of information that we will call **fragments**, each piece has been transformed according to a required level of protection such that it does not necessarily bear a sensible meaning for a human or even for a machine. This property certainly is at the very root of the idea of using and developing this mechanism to protect digital information. In this project, a fragment will be a unit of storage.

Moreover, a fragmentation system should easily comply with Kerckhoffs' second principle (see the original papers (in French) [Kerckhoff 1883] or the comprehensive history book on cryptography [Kahn 96]): it does not matter if a potential attacker knows how the fragmentation system works as long the location where fragments are stored and the way they should be decrypted and assembled is kept secret.

We call **F** the set (or collection) of all fragments relatively to a data the end user wants to protect. We denote by n = |**F**| the number of fragments. For terminology reasons, we propose to differentiate the notion of segment from the notion of fragment: a **segment** is a particular fragment separated from the rest of the information along natural lines of division (typically a paragraph in a text, a group of records, a table in a database, a column of data of a column-oriented database (see section 7.3), a group of objects, or a paragraph in a text); a **fragment** is a part, usually broken off. In general, a segment will have an understandable meaning for a human or a machine while a fragment is a more general piece of information (a sequence or a matrix of bits) possibly with little or no understandable meaning when considered without a context. Indeed, under these definitions, a segment must be seen as a special kind of fragment. Sometimes, the words chunk, share, or block can be found in the literature as synonyms for fragment. In this project, we will use the word **chunk** to define a first cut in clear of the original data before it is analyzed and protected, possibly aggregated for efficient storage purpose.

We call **map** the information describing where fragments are stored, how to access to them, how they can be decrypted, and how they must be re-assembled. Typically, the cardinality of **F**, file names containing fragments, theirs location, and access rights, as well as encryption keys or the order along which fragments must be stitched back together belong to the map. Sometimes, one of these aspects will be missing, being either considered as obvious by users or kept confidential outside of the digital world (e.g. a piece of paper in a safe, in the mind of the end user). A map can be implemented in multiple ways; it is supposed to be a small piece of information. Of course, a map can be itself fragmented and distributed between several users; it can also be hidden in selected fragments.

A public environment is almost by definition, particularly vulnerable. Any user can access to many servers loading persistent software that later have the ability to observe data in a stealth manner without being easily detected [Constantin 12]. We have to assume that computation or data is observed on at least one undetected machine. How to ensure that: *never all fragments are stored or even transit through the same machine* or in a somewhat weaker manner: *never enough fragments to provide meaning are stored or transit through the same machine*? If that was the case, fragmentation would be much weaker: only the order in which fragments pass by in front of an observer could still confuse him together with the fact that they belong to the same collection of fragments. To this regard, one can easily imagine building additional fake fragments (still belonging to **F**) as decoys. A good decoy would be extremely easy to discard for the authorized user and of course, utterly confusing for the attacker.

Fragmenting to achieve information protection can be considered as a simple idea; however, it is not so simple to efficiently design and develop a fragmentation system (including defragmentation of course) in order to comply with the motivations we just described and at the same time to avoid an important foreseeable overhead.

One key question lies in the way fragments are built so *the observation of one fragment is inconsequential*.

A simple strategy consists in separating critical information and to encrypt it from less sensitive information that can be let clear or plain. Beside criticality, size of fragments has to be taken into account: too small a size could lead to an unbearable overhead with a very large map, too large a size could create fragment bearing too vulnerable and large piece of information.

At the same time, it would be nice to fragment with enough redundancy to address an availability requirement such as: *the loss of one fragment is inconsequential*. This idea was already outlined in [Fabre…94] without much detail though. In [Shamir 79], this resilience requirement is supported with a notion of threshold scheme (more details in 7.1.1). We say that **F** has a **threshold** of k if and only if any subset of k fragments is needed to reconstruct the entire data while k-1 is not sufficient. In that case, *the loss of any n-k fragments is inconsequential*.

All these considerations clearly suggest that a naive partition and replication rarely will be an effective fragmentation regarding both data protection and resilience.

It must be stressed out that data nature (text, image, audio, video, 3D…) and format (html, pdf, bmp, jpeg…) can also influence the way fragmentation is performed. Data may be organized in an object oriented way. It can be stored in databases with records. Data can sometimes be segmented along these kinds of structures ([Fabre…94], [Ciriani…10]). More details are described in section 7.3.

### 3.2. Defragmentation

Defragmentation consists in reconstituting the information necessary to the end-user from fragments to be found in the storage area. Of course, the place where defragmentation is performed is a place of high vulnerability since this is where a meaningful piece of information is assembled then made clear for the end user. Defragmentation can be perceived with a triple challenge: Where are fragments requested by the user to be found and are they accessible? How many of them are needed? How should they be deciphered, stitched, and re-assembled together? Without a map as defined in 3.1, defragmentation can be as tangled as firstly finding many needles in many haystack followed by resolving a jigsaw puzzle, especially when fragments arrive in a random order. The smaller are the fragments, the more they are, the more defragmentation will be involved and complicated. It is also possible that only a few of them are needed to answer a specific user query. Therefore, some effort must be dedicated to design efficient defragmentation algorithms and methods, including their protection.

Moreover, it is important to include an integrity requirement to this task. Fragmentation can be bundled with data transformation (like encryption): the process of fragmenting defragmenting information must avoid arithmetic operations on real or floating point numbers with rounding possibilities to preserve integrity (this point will be discussed in details section 6.4.3.). Another operation that can produce integrity problems is transmission through communication channels with a given error rate. Fragmentation has to be designed with this kind of possible problem and avoid cascading effect.

We can consider defragmentation as the end of fragments life cycle: if after defragmenting and processing (i.e. modifying or updating) his information, a user wants to store again his information, we propose that he fragments anew his information creating a new brand family of fragments by changing a couple of parameters or seeds. By doing so, we would not have to trace and record which changes modify which fragment and where modified fragments have been stored and we would not have to verify and ensure that fragments are stored back to their original physical machine. These ideas involve resolving the questions of fragments clean up and deletion which can be tricky when thinking of backups. These questions are not addressed in the literature: a good number of papers describe one method of fragmentation; very few of them tackle the non-trivial question of defragmentation. We have not found any paper looking at the entire picture covering the entire information life cycle.

## 4. Technical challenges and high level requirements

### 4.1. Quality and cost effectiveness of data protection

Here, we consider four different dimensions: the volume or size of data to be protected, the level of protection the user wants to reach, the level of confidentiality or criticality of the data, the level of trust a machine is supposed to have and provide. Obviously, the more a data is confidential, the more the user wants to store it in a trustworthy place. Although they are different, there is a natural strong relationship between level of confidentiality and level of protection since an end-user will want to protect information in proportion of its level of confidentiality. We believe that a large amount of uncompressed data contains uneven levels of confidentiality, therefore, the need for protection and machine trustworthiness can gradually be set in proportion of the level of confidentiality. Figure 1 shows data represented by what we nicknamed a 'cone of confidentiality' to illustrate our thesis that in a large uncompressed amount of data, there is most of the time a large amount of data that can be considered as non-confidential and a relatively small amount of data that must be considered as confidential. This unproven thesis directly leads to consider multilevel security systems since they directly infer cost effective methods of storage on machines with various level of trustworthiness. Hybrid cloud architectures respond to similar tradeoff between cost effectiveness and confidentiality.

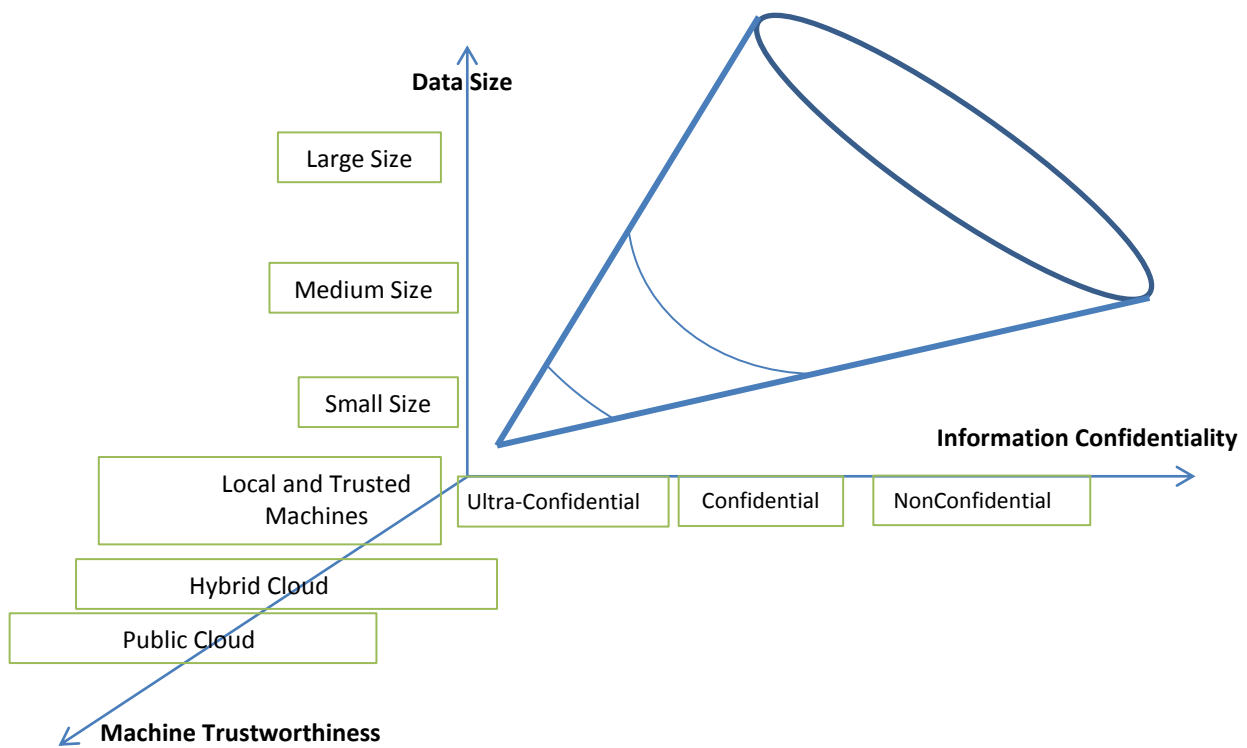

Figure 1 : 'Cone of confidentiality': Data confidentiality, machine trustworthiness, and data size

Assessing risks and defining a notion of *quality of data protection* is not an easy task especially with this concept of multilevel security. More than one user can be hesitating about what is sensitive from what is not. In section 6, we show how algorithms used in data compression can be used to separate information and support in automating this decision making. It can be addressed partially for instance, by comparison with classical encryption standard such as AES [AES 01] of the full information to be protected. Beside this kind of benchmark, statistical methods can be considered (see section 6.6), then more intensive security tests can be performed.

An attack can evolve over time and spread over several servers. An attack can be the result of a coalition of several attacks. The quality of data protection can vary with regard to the severity of the attack. This is not among the least elements that contribute in making this notion so difficult to master.

### 4.2. Performance and energy management

Fragmentation and defragmentation are resource demanding: it consumes extra CPU time and energy. It also often consumes more memory, more importantly it …fragments information among several physical storage devices since fragments should be stored in a discontinued and distributed fashion to achieve a strong level of protection.

Measuring performance particularly the overhead due to fragmentation and defragmentation is critical to understand how successfully this technique can potentially be deployed. To this regard, a user should accept to pay a price in overhead compared to dealing with his data with no protection. Fragmentation ought to be definitely faster than a full data encryption and must target performance a user can tolerate.

Parallelization in general, will obviously be the privileged way to obtain acceptable performance compared to a full data encryption; since data is fragmented, encryption (and decryption) of data can be performed in parallel on different servers or processors for different fragments. Size of fragments and strength of the encryption algorithm must be carefully chosen, implementation tradeoff has to be made since too many small fragments can create an unwanted additional and expensive overhead (a variation of Amdahl's law should apply here) and too large fragments can lead to modest to poor encryption performance.

When servers are geographically dispersed, it is important to consider latency. From a user point of view, latency is more acceptable during fragmentation than during defragmentation. Users may consider they are done as they hit the 'send' key of their keyboard and unless they wait for an acknowledgement, they do not care too much about the time taken for the fragments to reach their lieu of storage. On another hand, users won't like to wait too long for retrieving their information once they request it.

We will see in section 6 how the usage of a GPU as a special case of parallelization, can make selective encryption effective, hence usable [Qiu…14], [Qiu…15].

Energy consumption is also to be taken into consideration, especially with devices using batteries such as smartphones or networks of objects.

### 4.3. Data life cycle and data management

We envision data protection during its entire life cycle: data creation, version updates, backups, or archiving should be taken into account in our project. Updating data on a fragment basis can be attractive but complicated: the map would require careful corresponding updates and defragmenting could require a tool at the level of a configuration management system.

Data is aging. With time passing by, data sees its level of confidentiality varying: frequently decreasing until becoming obsolete; seldom increasing although this can happens for instance, for protecting the privacy of someone becoming important. Another example would be the discovery of a strong correlation between a highly confidential data and a data considered not confidential. This evolution could be particularly interesting to look at when considering long term storage or archiving. This evolution can become a difficulty for anonymyzation methods which let clear a number of records in database.

Data deletion on a fragment basis can pose similar problems as data updating.

To be complete, data duplication managed by File systems (which are constantly fragmenting and defragmenting files for memory management reasons) and more generally by Operating System layers (including temporary files, log files, caches,…), data backups, and archiving also have to be considered since they potentially can inadvertently regroup dispersed fragments on the same server or storage space. In any case, they have the ability to duplicate fragments and store them on physical machines out of the control of the protection system and the end-user, maybe right within the visibility range of our *uninvited observer*. This interesting case will not be studied in this report.

Another issue when dealing with databases is with services (e.g. searching) that can utilize data structures which could be centralized such as caching, indexing using hash tables, or dictionaries. Web management technology also can have an impact with tools like crawlers or ontology builders. It is important to analyze whether these important data structures overcome the barrier of fragmentation and betray or multiply data locations. More generally, data duplication creates additional exposure and should be considered.

### 4.4. Cloud Computing and Virtualization

Cloud computing providers commonly offer services at three different levels: SaaS (Software as a Service), PaaS (Platform as a Service), and IaaS (Infrastructure as a Service) through contracts. At the two first levels a customer will be offered Virtual Machines (VMs) for storing data and processing applications offered at the SaaS level. Virtualization mechanisms can prevent the end-user to store meaningful clusters of fragments in different physical machines by storing fragments in Virtual Machines without controlling where these VMs are assigned (which is after all, exactly what virtualization is all about: having the end-user not to have to care where is data are stored!). On another hand, it is to be noticed that most cloud providers allow assigning VM to specific datacenters. It is also possible to design an architecture involving several cloud providers as proposed first in [Aggarwal…09], then [Hudic…12] or [Bohli…13] for applications where latency is not in a problem for the end user or for data which have no privacy regulation constraints and for instance, can be exported in any country. Of course, this supposes that providers are not communicating with each other, that they are not collaborating for a powerful third party. An interesting case about multiple cloud is to consider hybrid cloud architectures where users can store their sensitive data in their own private cloud that is under their entire control. A side advantage against an attacker is that multiple providers can imply multiple software environments including multiple operating systems. This level of heterogeneity can pose additional difficulties and in any case, increases the cost to observe or attack stored information.

Another issue with virtualization has to do with memory consumption: encapsulating a single possibly small fragment per VM can be over expensive; a method will have to be designed to solve this issue.

Eventually, fragmentation should first be studied within use cases where virtualization can be avoided: at the IaaS level or within cloud management systems where virtualization can be avoided by users such as bare metal cloud, e.g. [TransLattice_Storm] (with a relatively costly kind of contract). Once this first level solved, one should consider virtualization.

### 4.5. Defragmentation avoidance

Clearly, fragmentation protects information only as long as data is …fragmented. The question is: can we process data without defragmention? This question is very similar to the one set for encryption which finds a potential answer with homomorphic encryption. We will look into this possibility in section 5.3.1.

#### 4.5.1. Partial defragmentation

In order to minimize risks and improve performance, it can be envisioned defragmenting just what is necessary to the end-user in a sort of lazy manner.

Partial defragmentation according to the subset of data needed ought to be investigated since a simple piece of information is generally sufficient to answer a user query. A practical solution for partial fragmentation is to have let the other fragments in clear as it is the case in [Bkakria…13] for instance (see section 7.3.2 for more details).

### *4.5.2. Total defragmentation avoidance*

It would be ultimately pivotal with regards to a wide adoption of these techniques to be able to avoid defragmenting information all together by being able to perform some computation directly on fragmented data. How to perform such distributed computation with reasonable performance? We may want to look at multiparty computation as introduced in [Yao 82] with a solution optimized in [Ioannidis…03] or searchable encryption as described in [Curtmola…11]. Of course, it would always remain to present computation results in clear to the end user; this step has to be discarded from what we have called 'total defragmentation avoidance.

# 5. Towards a complete Fragmentation, Encryption, Dispersion System (FEDS)

We call FEDS (for Fragmentation, Encryption, Dispersion System) a data protection system which includes the three phases: fragmentation, protection, and dispersion over a distributed environment. A 'complete' fragmentation system will include not only a fragmentation subsystem but should also include an effective defragmentation subsystem; it will have to cope with the technical challenges and high level requirements described in the former section. Figure 2 shows a possible workflow where fragmentation comes before encryption. This is to favor parallelization: fragmentation in the first phase of the process enables parallelizing encryption over as many fragments as CPU or threads are available. Any FEDS should carefully optimize its architecture for maximizing parallelization. Our notion of map is critical to obtain an efficient defragmentation; map management should not be forgotten in a FEDS design since it can be considered as the weak point of the system.

So far, we only found few systems dedicated on protecting large amount of data stored on large, powerful, and distributed systems integrating fragmentation with various methods of data protection. Section 7 describes and compares a handful of such academic and commercial systems.

## 5.1. Fragmentation System

Data can be of different nature and format: text, image, sound, or video; it can be stored in simple files or various types of databases, SQL databases as well as noSQL databases as HBase or Cassandra. This may influence the way to consider and analyze information and support producing fragments at different levels of protection.

Different environments and distribution of computation can be experimented with depending on the use case and its specific requirements under consideration. In this project, we used different environments; we particularly experimented with two different GPU architectures found on a desktop and a laptop. Then, we also used the Teralab platform (a tera-memory platform) for scalability studies; they are succinctly described in Annex: technical environment and access to the code. These different environments influence the definition of the best distribution of computation and justify to constantly evaluate the implementation to understand which software architecture fits best to a given hardware platform.

Designing a FEDS, it will be necessary to proceed carefully and stepwise, to experiment with scalability both in terms of size of information and size and number of fragments in order to understand how fragmentation and encryption method and implementation scale up or not, and eventually, test for security of course.

A complete fragmentation system will have to address most of our different high level requirements described earlier in section 2.

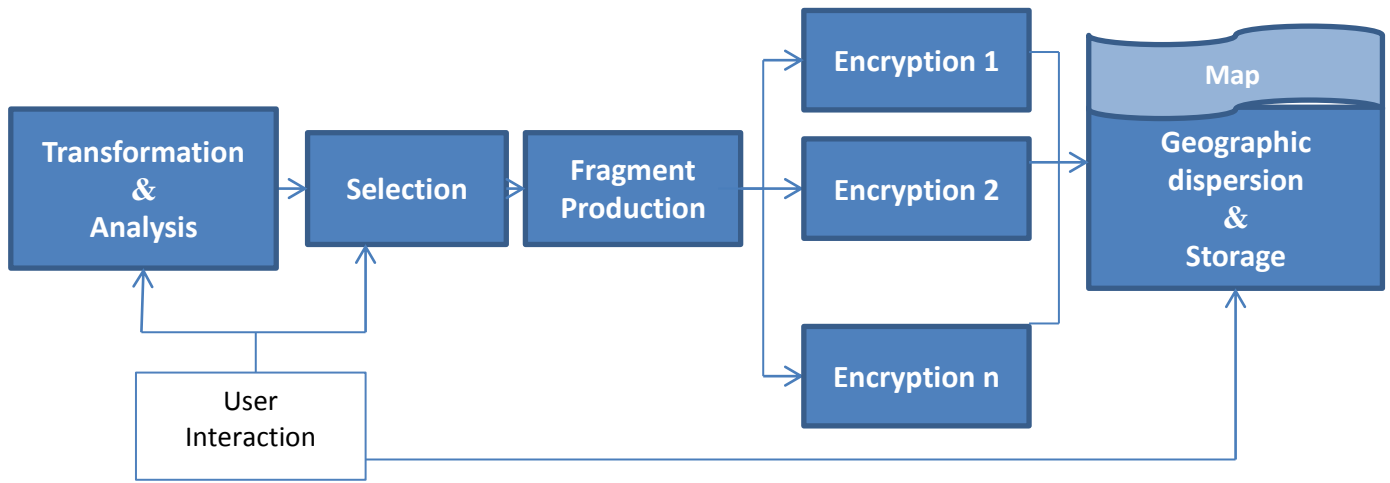

**Figure 2 :** *Basic workflow of a FEDS system*

Information candidate to be fragmented should be first parsed and possibly transformed for being easily and efficiently analyzed in term of level of confidentiality to determine how to separate critical data from non-critical data, to decide which parts must accordingly be strongly encrypted eventually stored in a private area, and which fragments can be let plain and possibly stored in a public area with as little risk as possible. Beside the levels of confidentiality, the analysis also determines how data could be fragmented in terms of size of fragments, degree of redundancy.

In our opinion, these two first steps should be permitted to be guided through user interaction to be effective: the user ought to know which information is critical which one is not, the user ought to know which level of protection he desires and take responsibility for it. He also can provide useful guidance about fragment size, strength of encryption, and location where to disperse and store which fragments.

However, we have already seen that in case of an image, automatic analysis could be performed to separate critical information from non-critical one [Qiu…14] or [Qiu…15]. The case where data format is a text is addressed in section 6.8 using transformation algorithms inspired from compression technologies. Similarly to what has been successfully done with image compression, several ideas drawn from text compression techniques seem promising and could be experimented with.

Fragmentation can be performed in several successive steps. For instance, in [Fray…86], files are divided in blocks then in fragments.

For each fragment i, an encryption algorithm i is chosen; it can be depending on the size of the fragment (i.e. to comply with performance constraints) and the criticality of the information it contains. This encryption algorithm can be the identity if it is decided to let the fragment clear. This is one fundamental advantage of fragmentation over full encryption:

fragmentation offers the capability to customize the level of protection, the strength of the encryption algorithm with respect to the need expressed by the end-user or criticality of the piece of information to protect; opening an interesting opportunity of optimization in term of cost (performance, energy, memory). A fragment can be used to encrypt in some way another fragment. This idea can be used to address key management.

It will be important at each step of the way to benchmark and compare results against the best known implementations of encryption algorithms used without fragmentation, tuning parameters to improve performance.

It is well understood that progress in technology, progress in encryption implementation efficiency are continuous (see for instance, Vampire Lab benchmarks). This can influence the degree of parallelization of encryption or fragmentation. It can change the optimal size of fragments. This remark constitutes a second justification for the selection step to be flexible.

The last step is to decide on which machine to store which fragment. Ideally, it is important to avoid storing fragments which, when gathered, reconstitute a critical piece of information. It is also clearly recommended to store replicated fragments on different physical machines. At the same time, it is also important to minimize latency. This implies having the knowledge of the topology of distributed machines at disposal.

### 5.2. Defragmentation System

Then, it will be important to design and evaluate different fragmentation and defragmentation methods in particular addressing few of the technical challenges described earlier paying particular attention to performance and analyzing in details causes of overhead and quality of data protection, for instance, by tuning size and number of fragments.

We then propose to understand how to dose encryption and fragmentation to reach a well-balanced solution addressing at the same time performance, reliability, security and privacy.

Each time, a protocol is derived, it will be necessary to understand its complexity and to validate, possibly prove its properties and behavior.

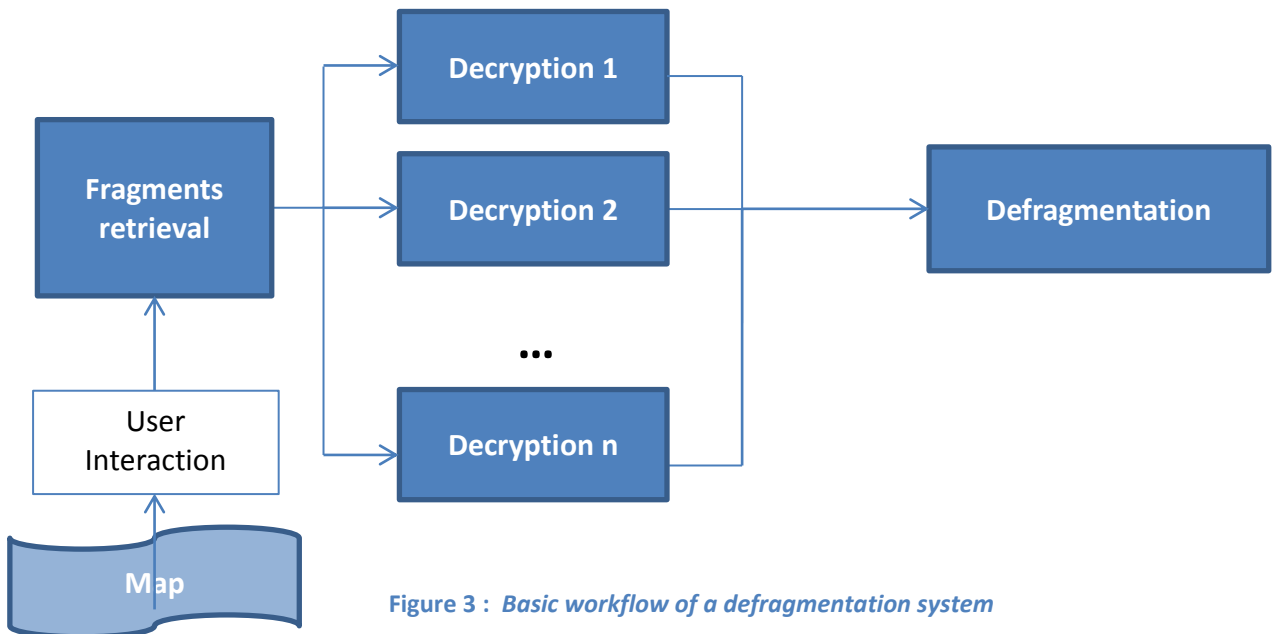

Figure 3 : *Basic workflow of a defragmentation system*

Fragment retrieval must be fast; this probably is the step where end-users will feel latency the most acutely and with little tolerance. The user interacts with the FEDS system by selecting a piece of information (possibly by query) not necessarily its whole. This step must be aware of where fragments are stored (using the map) and which fragments will be useful to the selection of the user.

Figure 3 is showing key elements of the defragmentation workflow; indeed, each decryption algorithm i corresponds to the encryption algorithm i used during the fragmentation. Again, it will use the map to access to the necessary key $k_i$.

Defragmentation can be pretty involved especially when fragments have been retrieved in a random order, when fragments overlap with each other, and when decoys have been produced during fragmentation and must now be discarded. This step can easily be parallelized where each machine processes and defragment a sub-group of fragments at a time. By doing so, some of the machines used for defragmenting can very well be in a non-trusted area.

At last, it should not be forgotten to perform the inverse of the transformation used in the first steps of the fragmentation workflow.

## 5.3. FEDS and data processing

Protecting data during its processing is utterly difficult and efficiency of any solution largely remains an open question in basic research since it is indeed a very desirable property [Rivest…78]. We consider two possible venues: the first one consist in processing data that remain encrypted, the second is about processing data while they stay fragmented to avoid defragmentation as much as possible since it is evidently a phase of high vulnerability.

### *5.3.1. Processing encrypted data*

Homomorphic encryption ([Gentry 09b]) and searchable encryption [Curtmola…11] are both attractive techniques since they both allow operating (arithmetic and logical operations for homomorphic encryption; search of a character string for searchable encryption) directly on encrypted data without the need for decrypting them. The stumbling block is that these methods, despite recent progress, are extremely complex and expensive in both memory and execution time. Using them for protecting massive amount of data does not seem at first, very reasonable. Fully Homomorphic Encryption (FHE) is a concept first envisioned circa 1978 by [Rivest… 78] and then reworked and developed in [Gentry 09b]. This concept can be described through this simple question by C. Gentry: "Is there a way to delegate processing of your data without giving away access to it? We immediately understand the value proposition of such encryption algorithms even before considering outsourcing or public cloud computing since it is about performing computation with encrypted data in perfect security. The trustworthiness question in cloud computing has been discussed for years and today, there is still no perfect solution. FHE could very well be this 'perfect solution' only once proven efficient from a performance point of view.

Figure 4 shows how FHE can be used within a public cloud server to compute the value of a function *f* for a data *Data*: the user Alice sends the encrypted *Data* (with a key *Key*) and the function *f* to the cloud and will receive back an encrypted *f(Data)* (with a key *Key'*). The homomorphic property of FHE allows the cloud performing the computation of *f(Data)* without knowing or accessing to *Data* or *f(Data)*. Alice will be able to decrypt *f(Data)*.

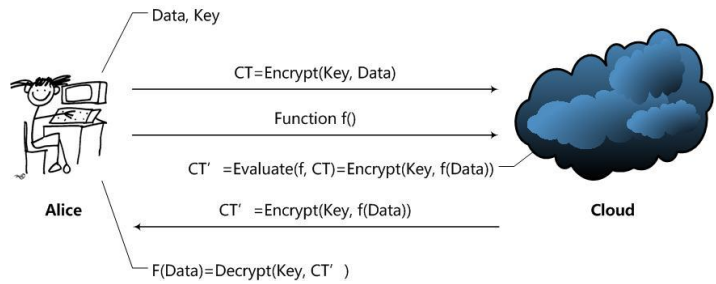

**Figure 4 How FHE works in a real scenario**

Other encryption algorithms are known for having partial homomorphic property[3]. For instance, RSA is homomorphic with regards to multiplication: If the RSA public key is modulus *m* and exponent *e*, then the encryption of a message *x* is given by:
$$\mathcal{E}(x) = x^e \bmod m.$$

The homomorphic property for the multiplication is then:
$$\mathcal{E}(x_1) \cdot \mathcal{E}(x_2) = x_1^e x_2^e \bmod m = (x_1 x_2)^e \bmod m = \mathcal{E}(x_1 \cdot x_2),$$

### *Related work*

The FHE was built on math basis that can prove it is secure. Since in 2009, FHE based on ideal lattice is introduced by [Gentry 09a], three main branches of FHE schemes have been developed: lattice-based, integer-based and learning-with-errors (LWE) or ring-learning-with-errors (RLWE) based encryption.

The main focus of the theoretical cryptographic research community is currently on LWE and RLWE based FHE. LWE was introduced by Regev, and has been shown to be as hard as the worst case lattice problems. This problem has been extended to work over rings, and this extension increases the efficiency of LWE.

Integer based schemes were introduced by van Dijk et al. as a theoretically simpler alternative to lattice based schemes and have been further developed to offer similar performance to existing lattice based schemes.

Despite different math basis have different performance, none of them is efficient enough for actual applications. For example, key generation in Gentry and Halevi's lattice based scheme takes from 2.5 seconds to 2.2 hours. And for the evaluation step, a recent research by Gentry et al. shows a homomorphic evaluation of AES requires 36 hours which is actually incredibly slow compared with hundreds MB/s with AES on a modern PC's CPU.

Another important limitation of FHE is with the memory usage. FHE generates very large cipher text and uses very large public keys to guarantee adequate security to prevent against possible lattice-based attacks. Gentry and Halevi's FHE scheme uses public key sizes ranging from 17 MB to 2.25 GB.

Current research is aiming at improving performance of FHE either by focusing on new basic mathematics results to reduce computation complexity or by implementing the existing FHE algorithms on different hardware (GPU or nanotechnology). New algorithms are expected to provide with an actual breakthrough in term of performance; however, on another hand, hardware progression is relatively limited with regards to the need for deploying FHE.

### *Performance study*

---

[3] see Wikipedia https://en.wikipedia.org/wiki/Homomorphic_encryption

In this section, we provide current research results about performance of the existing algorithms and their implementations. We are adding our own implementation for comparison. As we mentioned earlier, theoretical breakthrough of algorithm may bring a revolution in term of acceptance of FHE, this may need many years of work. In the meantime, it is interesting to search for possible optimized solutions including by using existing powerful hardware to determine whether FHE is ever usable. Although many research articles have claimed the performance of FHE are slow or far from application, it seems important to characterize how slow FHE really is. Performance of the underlying crypto-primitives such as modular reduction and large multiplication are required in many of the FHE schemes. Actually, they are critical these operations could be significantly improved through the use of GPU, FPGA, or Asic technology.

The first GPU implementation of a FHE scheme was presented by Wang et al. in 2012. The authors implemented the small parameter size version of Gentry and Halevi's lattice-based FHE scheme on an NVIDIA C2050 GPU using the FFT algorithm, achieving speed up factors of 7.68, 7.4 and 6.59 for encryption, decryption and the recryption operations, respectively. The FFT algorithm was used to target the bottleneck of this lattice-based scheme, namely the modular multiplication of very large numbers.

An overview of previous FHE implementations on various platforms is presented in `Table 1` taken from [Doröz…13] where we added two additional implementations of our own, the first one on a CPU, the second one using our GPGPU. Clearly, since the platforms vary greatly according to available memory, clock speed, area/price of the hardware a side-by-side comparison is difficult and therefore this information is only meant to give an idea of what is achievable on various platforms.

Much of the development so far focused on the Gentry-Halevi FHE, which intrinsically works with very large integers (million bit range). Therefore, a good number of works focused on developing FFT/NTT (Number Therotic Transform) based large integer multipliers. Currently, the only full-fledged (with bootstrapping) FHE hardware implementation is the one reported by [Doröz…15], which also implements the Gentry-Halevi FHE. At this time, there is a lack of hardware implementations of the more recently proposed FHE schemes, i.e. Coron et al.'s FHE schemes, BGV-style FHE schemes and NTRU based FHE.

| Designs | Schemes | Platforms | Performance |
|---|---|---|---|
| **CPU Implementations** | | | |
| AES | BGV-FHE | 2.0 GHz Intel Xeon | 5 mins/AES block |
| AES | NTRU-FHE | 2.9 GHz Intel Xeon | 55 sec/AES block |
| Full FHE | NTRU-FHE | 2.1 GHz Intel Xeon | 275 sec/bootstrap |
| Full FHE (our test) | BGV-FHE | 3.0 GHz Intel I7 | 3-5 mins/bootstrap |
| **GPU Implementations** | | | |
| NTT mul/reduction | GH-FHE | Nvidia C 250 | 0.765 ms |
| NTT | GH-FHE | Nvidia GTX 690 | 0.583 ms |
| AES | NTRU-FHE | Nvidia GTX 680 | 7 sec/AES block |
| NTT mul (our test) | GH-FHE | Nvidia GTX 780 | 0.81 ms |
| **FPGA Implementations** | | | |
| NTT transform | GH-FHE | Stratix V FPGA | 0.125ms |
| NTT mod/enc | CMNT-FHE | Xilinx Vitrex-7 FPGA | 13 ms/enc |
| AES | NTRU-FHE | Xilinx Virtex-7 FPGA | 0.44 sec/block |
| **ASIC Implementations** | | | |
| NTT mod | GH-FHE | 90 nm TSMC | 2.09 sec |
| Full FHE | GH-FHE | 90 nm TSMC | 3.1 sec/recrypt |

Table 1 Performance study of FHE taken from [Doroz…15] where we added our own tests framed in beige

*Conclusion*

Results for different FHE algorithms and for limited evaluation functions (AES-128 bit here) were presented in Table 1. As in the European H2020 ECRYPT project [Archer…15b], we can conclude that FHE is still today far from real application. But here, thanks to this table, we can quantify the issue. The AES block is processed in about 1-5 mins on an Intel Xeon CPU which is the type of CPU currently used in workstations. A good GPU (Nvidia GTX 690) could help reducing this processing to about 7 seconds. However, considering that AES is processed at a hundreds MB/s on PC's CPU, which equals almost 1 million blocks processed per second, the performance of FHE-AES is far too slow to be considered usable. Even if hardware platforms are upgraded, even if performance of the FHE-AES algorithm is improved one thousand times, FHE would be still too slow to be vastly deployed.

The merit of Table 1 is not only to confirm the performance issue with FHE but also to tell us the magnitude of needed progression before deploying FHE. Our own code is on par with current publications. One possibility would be to use partial homomorphic encryption (PHE) or somewhat homomorphic encryption (SWHE) but in any case, their usage would stay very limited to niche applications. Another possibility would be to go towards fundamental mathematical research and devise new multiplication algorithms like in [Rambaud 15] despite high difficulties to progress in this kind of direction.

### 5.3.2. Fragmenting the computation

It seems more promising and in perfect coherence with our concept of fragmentation, to look at multiparty computation ([Yao 82]) and investigate the scalability question in the light of new hardware powerful platforms such as GPUs, Teralab, or multicore machines. The European H2020 ECRYPT project [Archer…15b] is suggesting this research direction.

# 6. Selective fragmentation, selective encryption

## 6.1. Introduction to selective encryption

Traditional standard encryption systems are commonly used to protect data (e.g. DES, 3DES, or its successor AES, etc.). However, it is argued (for instance, in [Massoudi…08a]) that these encryption systems, which have been originally developed for text data are less suitable for securing images mainly because they consist in putting in the whole image data into a standard encryption system without considering its specific nature. One issue [Krikor…09] is that all symbols in the content are of equal importance are argued non optimal for securing images. Another issue which will be addressed in this report is with performance. Full encryption algorithms can be time consuming once the end-user requires speed with regard to the protecting process while disposing of limited calculation resources environment like the ones available in a laptop. We could observe by experimentation that encryption and decryption processing of a bitmap image on a laptop causes non-negligible overhead. It is therefore important to understand how to speed up this processing. Many methods have been proposed and developed in the past years (for instance [Ziedan…03]). Using weaker encryption algorithms or sequential implementation of selective encryption either have a lower security level or have poor CPU performance.

Other works are proposing other methods for securing images, in particular methods called selective encryptions (SE) which are the focus of this work. SE consists in applying encryption to a subset of the original content with or without a preprocessing step. The general approach described in Figure 5 : General concept of selective encryption.Figure 5 is to separate the image content into two parts or fragments. The first fragment is to be confidential and will be encrypted. The main goal of SE methods is to reduce the amount of data to be encrypted and take as little storage as possible while achieving a required level of security. The tradeoff is to make the confidential fragment as small as possible in order to reduce processing time while keeping the image secure enough to comply with the requirements of a given specific use case. It is the task of the preprocessing step (in Figure 2) to resolve this tradeoff and separate the image in two fragments.

In this first experiment, information theory is used to separate and reduce the amount of data that needs to be protected and encrypted.

In order to understand which part of the image will be encrypted, we have to transform the image. First, we used a Fast Fourier Transform algorithm, then a Discrete Cosine Transform algorithm (DTC) to get the frequency domain distribution of the data. We can then easily do fragmentation by judging which part is more important. Then different encryption methods to different parts of data can be applied according to their level of criticality. It turns out that the DCT is a reasonable balance of optimality of the input decorrelation (approaching the Karhunen-Loève transform) and the computational complexity.

The second fragment is intended to be public and unencrypted as such, this fragment should not be sufficient to reveal or restore the full information. Moreover, it is intended to take most of storage space.

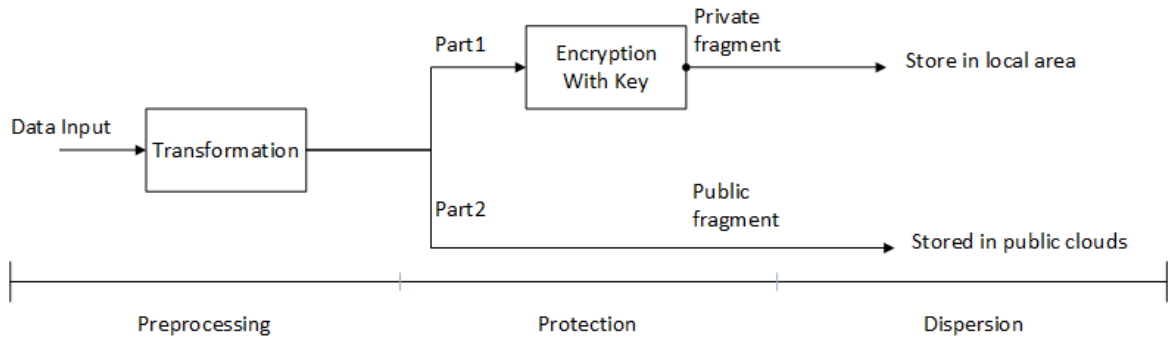

Figure 5 : General concept of selective encryption.

For uncompressed images like bitmaps, the most important visual characteristics of an image are to be found in the low frequencies while details are situated in the higher frequencies. Studies on HVS (Human Visual System) have confirmed that human are more sensitive to lower frequencies than to higher ones [Puech…05]. This is why most SEs select and encrypt low frequencies in the confidential fragment rather than high frequencies. However, it is known that image sharp details reside in high frequencies; this means that sometimes, the public fragment can unveil information. We will discuss this point later since it led us to define a new design for SE with a strong level of protection.

There exist methods to selectively encrypt values in frequency domain [Krikor…09] to protect bitmap images. However, although providing a good level of protection, these methods actually suffer from performance issues, a major impediment which made SE difficult to use. These performance issues will be extensively addressed in this report.

We are interested in image protection within limited calculation resources environments like desktops, laptops, or mobile devices such as latest tablets or even smartphones. In this work, bitmap files are initially stored on a laptop. Two new methods of selective encryption in the frequency domain are presented, called first level of protection and strong level of protection. They both use a GPU (Graphic Process Unit) to provide with the necessary acceleration, shifting the heavy computation burden from CPU to GPU. They aim at speeding up SE method by using calculation resources of both CPU and GPU available on a laptop.

This report is including results of our paper presented at ISM'14 [Qiu…14] and describes a novel strong level of protection for selective encryption published in [Qiu…15].

In section 6.2, we discuss previous works on selective encryption methods and their performance issue. In section 6.3, we present the two designs of our method in particular, introducing a new design which provides a strong level of protection. In section 6.4, the usage of GPU for the computation of the Discrete Cosine Transform (DCT) is discussed. Consideration on the accuracy of the computation is also addressed since it directly impacts the quality of the image reconstruction. In section 6.5, we describe parallel implementations and extensively discuss results by considering two different GPU architectures. In section 6.6, statistical calculation is processed to analyze the level of protection of our methods. We conclude in section 6.7 by suggesting how selective encryption can be used to safely store public fragments in public storage systems, then, we discuss a last hardware architecture point that influences implementation before hinting on future work.

## 6.2. Related work & Performance issues

### 6.2.1. Related work

SE schemes have been described and discussed in several previous works [Uhl…05], [Khashan…14], or [Massoudi…08a]. In their book, [Uhl…05] are surveying image and video encryption paying little attention to selective encryption. It is suggested that this method can be used for light encryption. The most likely reason certainly is about performance. According to [Massoudi…08a] the ratio between the respective sizes of public and confidential fragments depends on the use case and its level of required protection. Various kinds of methods are adapted to protect different image formats (JPEG and so on) or different image contents.

In [Massoudi…08b] selected bits of each byte are protected considering the image as a binary file. This kind of method works well for JPEG image protection because the JPEG image files contain DCT coefficients after quantization and Huffman coding. These coefficients vary in a large range of values: the DC coefficient and the first low frequency coefficients are much larger than the remaining frequency coefficients. These differences in coefficient values provide the basis for the selective encryption methods by selecting and protecting the most important values. However this method must be refined for uncompressed bitmaps since the selected confidential fragment could be too large. For instance, in [Munir 10], four most significant bits (MSBs) of each pixel which represent 50% of the original data volume are selected for encryption.

There exist methods to directly protect images without a preprocessing step like in [Zhang…13], selectively encrypting some areas by using scalable shape context to locate the important character in a cartoon image. This method aims at protecting only the most valuable area in the image. The remaining public part will be left plain therefore vulnerable. Moreover, the first step consisting in separating the two parts is time consuming and affects the overall performance.

Images can be protected in the frequency domain by encrypting some selected coefficients after processing the DCT algorithm. In [Puech…05] a method applied to medical images consists of encrypting the sole DC coefficients. The DC coefficients in the DCT represent the average intensity of the DCT block which is critical from an energy viewpoint. As shown in [Puech…05], protecting the DC value can highly degrade the visual quality of image or even make an image totally unreadable. However, as pointed before [Uehara…06], DC coefficients can be recovered from the remaining coefficients. So protecting an uncompressed image in the frequency domain should not only protect the DC values but also some AC values as well like in [Krikor…09] and [Yuen…11].

According to our research, protecting the DC coefficient and the first five AC coefficients in a DCT 8×8 block can efficiently protect images in most cases as it degrades the remaining public fragment into an almost totally unreadable and unrecoverable image. However, some images may contain many sharp edges which are transformed into some significant high frequency values. If we just use the remaining coefficients to do a reverse DCT (ICDT), some parts of these images may still be restorable to a readable image. This is the reason why the number of AC coefficients belonging to the confidential fragment ought to be tunable along the desired level of protection.

Moreover, performing DCT in the preprocessing step in order to separate uncompressed images like bitmaps into two parts can suffer from too an expensive performance overhead which can make SE even slower than a full image encryption. It is therefore important to benchmark any SE implementation against a full image cipher like AES.

### 6.2.2. Comparing AES and DCT implementations

Very few papers discuss and show performance evaluation of SE algorithms implementation like in [Khashan…14] where full image encryption with AES is compared to author's SE implementation. However, it is important to benchmark SE against a full image encryption with cipher like AES in order to understand their actual potential usability. The speed of standard block encryption algorithm today is actually very fast, e.g. the AES 128-bit can reach more than 200MiB/s on a PC's CPU [Dai 09]. Assuming AES 128-bit is the encryption algorithm used after a preprocessing of an SE algorithm, we can conclude that a SE implementation will have acceptable performance from a user point of view only if the select step is faster than AES 128-bit.

Moreover, as SE algorithms are based on the versatile nature of multimedia contents, a general rule to judge whether a SE algorithm provides with a good level of protection or not is difficult to define since we have seen that this kind of algorithm efficiency depends on the nature of the multimedia content which can be very versatile. We consider a SE algorithm as 'usable' if this algorithm meets both a suitable level of security with regard to the needs for the special use case it is supposed to address and a level of performance comparable or better than a full encryption algorithm (in this report, we use AES 128-bit as the standard encryption). Based on this definition, most of the SE algorithms using DCT 8×8 in the literature are actually 'unusable' as DCT 8×8 algorithm is not faster than AES running on the same CPU. We use two different CPUs to test this assumption: the speed of AES (in two modes: Cipher-Block Chaining (CBC) and Cipher Feedback (CFB)) is compared to the speed of DCT 8×8. Table 2 shows that DCT 8x8 is around 45% slower than AES.

| Intel CPU | AES/CBC 128-bit | AES/CFB 128-bit | DCT 8×8 |
|---|---|---|---|
| I7-3630QM | 374 MiB/s | 362 MiB/s | 203 MiB/s |
| I7-4770K | 494MiB/s | 480 MiB/s | 267 MiB/s |

Table 2 : Benchmark of AES 128-bit and DCT 8×8 on current CPUs.

In order to speed up computation of DCT and make SE 'usable', we proposed in [Qiu…14] to use a GPU (Graphic Process Unit) as an accelerator. This kind of processor exists today in many laptops, desktops, tablets or even smartphones. GPUs offer hardware-level parallelism with hundreds or thousands of slim cores [Owens…08], [Han…10]. Their data-parallel execution model fits nicely with image transformation between time domain and frequency domain like DCT. AGPU acceleration for DCT 8×8 (DCT by block 8×8) is already shown in [Patel…09] or [Obukhov…08], however, not in the context of SE. Another reason for using GPU is their rapid density improvement which goes faster than usual CPU improvement [Han…10].

### 6.3. Designs

Two designs of selective encryption in the frequency domain are described: a first level of protection (already described in [Qiu…14]) when speed is of the essence, and a more complex strong level of protection when a more global protection of the image is required. On the one hand, our design proves that encrypting coefficients in frequency domain can efficiently protect images; on the other hand, this work shows that the data-parallel execution model of GPU fits nicely with the preprocessing step, DCT 8×8 which transforms images from spatial domain to frequency domain block by block. This feature will makes GPU provide with such a performance gain for selective encryption using DCT 8×8.

The first level of protection is separating the vital piece of an image in a similar way than in [Krikor…09]. However, we found that the public fragment still may contain some visual characters that could result in unveiling important information. This issue has been the starting point to designing a strong level of protection following the same general concept as in Figure 5 and preserving similar performance.

### 6.3.1. First level of protection

In our previous design [Qiu…14], we refined the "Preprocessing" step with more details (see Figure 6). First, the input data will be preprocessed using DCT 8×8. Then the results of the DCT 8×8 which are the coefficients in the frequency domain will be fragmented into two parts according to the selection ratio with respect to the protection level which is desired. The encryption system will be used for the confidential part, Part 1. Part 2, the public one, is let plain.

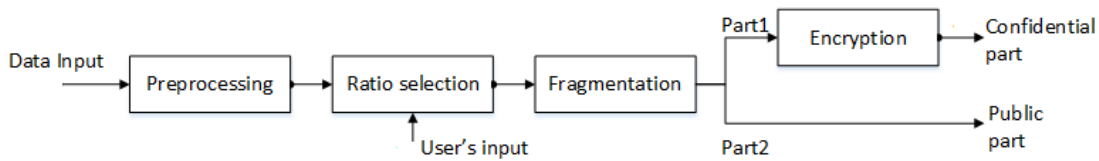

Figure 6 : Our design method for a first level of protection where part 2 is let plain.

This method significantly reduces the need for the data to be fully encrypted and improve the output performance. The fragmentation step possesses selection ratio done in the frequency domain to increase or decrease the confidential part to be encrypted allowing the user to increase or decrease the desired level of protection. The fact that Part1 is small is based on information theory saying that: the distribution of the information energy is not uniform in the frequency domain. Most of the energy concentrates in the low frequency part which takes a small percentage of storage (Part1 is usually less than 10% of the total footprint). The ratio selection can be set by user input. However, the coefficients [0,0], [0,1], [1,0], [2,0], [1,1], [0,2] of the DCT 8×8 results cited from [Krikor…09] constitute the default selection and is recommended from experience. Selecting fewer coefficients for the private part would let a public part revealing too much. According to a non-uniformly distribution property of the energy contained in an image signal, the DC coefficient of the DCT result and another 5 AC coefficients are usually taking over 95% of the energy. This was confirmed by testing multiple images.

This protection method will erase most important visual characters from an image (like people's face). It is recommended if for instance, the user's target is to protect attackers from knowing who is in the image. More generally, this first level of protection is good for soft encryption when high performance is required at the same time. In Figure 7, we select two original images (a) and protected results (b).

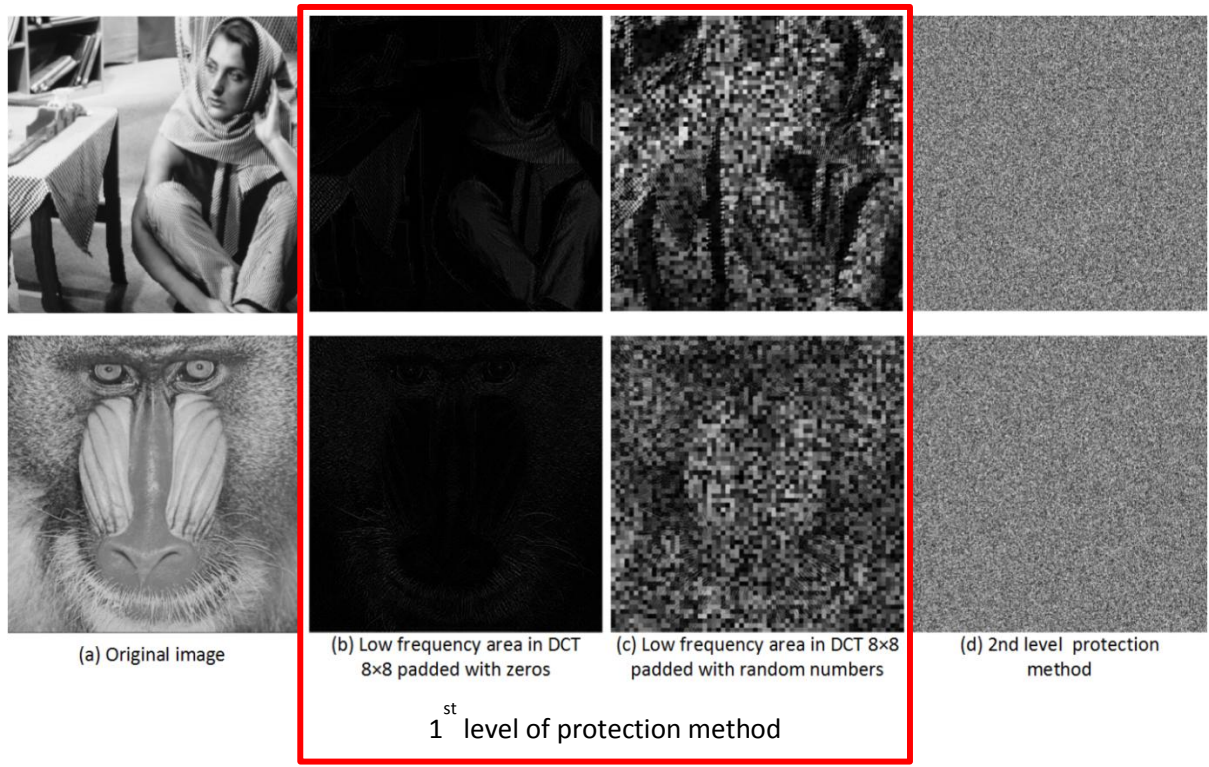

Figure 7 : *Original image (a), images where the low frequency domain is padded with zeros (b) then with random numbers (c), and strong level of protection (d)*

### 6.3.2. Strong level of protection

As said in introduction section 5, the level of protection depends on the nature of the content itself. High frequency coefficients sometimes may reveal some piece of the original image, since an image may contain some sharp edges or clear details. If we put random values in the low frequency domain and DC value, details like sharp edges of the original image will be enhanced (random numbers will enhance the correlation of the remaining DCT coefficients) and can reveal details as shown in Figure 7 (c).

For some user cases, this kind of protection would not meet requirements. For example, if the content of an image is sensitive and requires a high level of protection, the high frequency coefficients need to be protected as well in order to prevent any visual elements to be unveiled. What must be noticed here is that performance for the whole selective encryption would indeed become worse than the full encryption if the same encryption method had to be used for the high frequency coefficients.

Hence, the slightly more complex design in Figure 8 where Part 1 which contains selected low frequencies is used to generate a key to build up a light and fast protection for the high frequency part. In this work, the key generation step uses the SHA-512 function [NIST 02] to get a unique fixed-length string (512 bits long) from the 6 selected coefficients (Part 1 in Figure 8). The SHA-512 function has a feature that can generate two different and unpredicted results even only one bit of the input string is different. This feature will guarantee that the 512 bit string cannot be predicted even adjacent 8×8 blocks of an image can be very similar. Because the block we processed is 8×8, the reverse DCT result of the rest DCT coefficients padded with zero contains 64 pixels storing in byte which is exactly 512 bits. The XOR step can protect every pixel block by block.

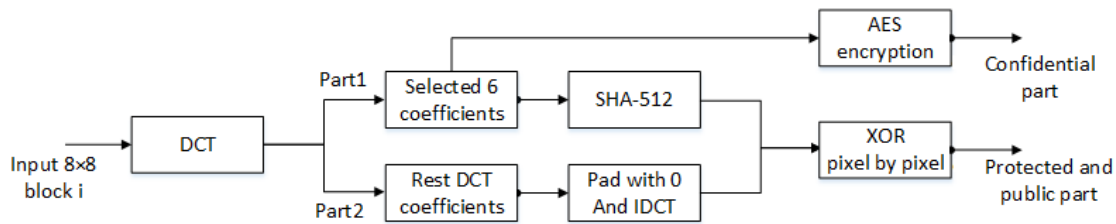

Figure 8 : Design to enhance the protection level.

As shown in Figure 7(d), the XOR pixel by pixel will thoroughly blur any visual characters and generates an almost uniform grey image. Section 6.6 will discuss this point from a statistical point of view.

## 6.4. Two key implementation choices

Before going further, we feel of paramount importance to focus in this section on DCT implementation, since, as seen in Table 2, AES implementations on a CPU is much faster than DCT. Here, the idea is to use a general purpose GPU for processing DCT. We also verify in this section that our implementation will be lossless (affecting the quality of the image reconstruction) and that the increase in memory due to this processing is particularly reasonable.

### 6.4.1. GPU architectures

Commodity graphics processing units (GPUs) have rapidly evolved to become high performance accelerators for computing. Over the past few years, the GPU has evolved from a fixed-function special-purpose processor into a full-fledged parallel programmable processor with additional fixed-function special-purpose functionality [Owens…08]. The modern GPU is not only a powerful graphics engine but also designed as a general purpose engine with a large number of processor cores [Ryoo…08].

As indicated in [Ryoo…08], the NVIDIA CUDA programming model [Nvidia 15] was created as an inexpensive, highly parallel system available to a continuously larger community of application developers. In this model, the system consists of a host that is a traditional CPU and one or more computing devices that are massively data-parallel coprocessors. Each CUDA device processor supports the Single-Program Multiple-Data (SPMD) model [Atallah…10], widely available in parallel processing systems, where all concurrent threads are based on the same code, although they may not follow exactly the same path of execution. All threads share the same global address space.

Let us notice that according to [Gregg…11], the calculation ability of GPUs varies a lot due to their architecture and configuration differences (particularly in terms of numbers of cores). In their paper, they show that the Geforce GTX 480 card runs a sort application more than 10 times faster than a 330M card.

This is why we used two different GPUs to implement and evaluate our design. The first one is an usual laptop GPU Nvidia Nvs 5200M with a compute capability version 2.1 and the second one is a high performance desktop GPU Nvidia Geforce GTX 780 with a compute capability version 3.5 which provides the board with a more efficient memory management. In Table 3, we compare the two GPUs used in our implementations with two of the GPUs used by [Gregg…11] (Geforce GTX 480, a desktop GPU and 330M, an old laptop GPU). We can observe the progression particularly in terms of number of cores. We will come back on this point in our conclusion in section 6.7.

| GPU Type | CUDA Cores | Memory (MB) | Clock (MHz) | Memory Width |
|---|---|---|---|---|
| Nvs 330M | 48 | 256 | 1265 | 128-bit |
| Nvs 5200M | 96 | 1024 | 1344 | 64-bit |
| Geforce GTX 480 | 480 | 1024 | 1401 | 320-bit |
| Geforce GTX 780 | 2304 | 3072 | 941 | 384-bit |

Table 3 : Detail information of GPUs used in our work and by Gregg & Hazelwood (2011).

### 6.4.2. DCT acceleration on GPU

The DCT is a mathematical transformation first introduced in [Ahmed…74] that takes a signal and transforms it from spatial domain into frequency domain. We used the DCT optimized algorithm from the Nvidia technical report [Obukhov…08]. According to this report, an important feature called separability is used to make 2D DCT calculated by multiplication of vertically oriented 1D basis with their horizontal representations which fits well with the GPU parallelized hardware architecture.

According to [Patel…09], the DCT 8×8 are implemented by using matrix multiplication which can be easily used by GPU acceleration. In [Huynh-Thu…08], it is reported that DCT 8×8 computation can be accelerated by a factor 10 to 25. In Table 4, we compare the acceleration for DCT 8x8 on our own two computers, a laptop with the 5200M GPU, an Intel I-7 3630QM 2.4GHz CPU and a desktop with the GTX 780 GPU, an Intel I7-4770K 3.5GHz CPU. This time, we can observe that DCT 8×8 is accelerated by a factor 10 to 75 with small variations when the image size varies. This shows how fast GPU architectures are progressing in comparison with CPU.

| Image size (in pixel) | 1024×768 | 1600×1200 | 3240×2592 | 4800×4800 |
|---|---|---|---|---|
| Laptop CPU time | 3.78ms | 9.24ms | 39.4ms | 108.4ms |
| Laptop GPU time | 0.41ms | 0.79ms | 3.67ms | 9.98ms |
| Performance gain | 9.2 | 11.7 | 10.7 | 10.8 |
| Desktop CPU time | 2.88ms | 7.0ms | 29.9ms | 82.4ms |
| Desktop GPU time | 0.04ms | 0.09ms | 0.41ms | 1.12ms |
| Performance gain | 72 | 77.8 | 72.3 | 73.6 |

Table 4 : DCT 8×8 acceleration for CPU and GPU along various image sizes

From this test, we see that CPUs on a laptop and desktop computer are not that different as the performance of CPUs mainly account on the factors like main frequency and caches. For the same generation of Intel CPU, the performance is actually quite similar compared with the huge difference of the GPU performance which directly rely on the amount of the CUDA cores (could vary from 100 to 2300). Also, DCT 8×8 on GPU can run more than 70 times faster on the desktop than on the laptop. This kind of acceleration for DCT 8×8 greatly changes and broadens the scope of selective encryption applications for bitmap images.

### 6.4.3. Storage space and numeric precision

There is a classic tradeoff between the memory occupation (in terms of both footprint and storage space) and numeric precision when it comes to handling floating point numbers with high precision. A similar tradeoff occurs with DCT computation. Each pixel in bitmap files is usually stored as an 8-bit integer (two more bytes are used for 'Highcolor' and two additional bytes are used for 'Truecolor'). During the DCT computation, these integers are transformed into floating point 32 bits numbers increasing the footprint by a factor 4. At the end of the computation, numbers are turned back to integer and this is repeated during the computation of IDCT. These changes of numeric data type can involve truncation and generate image distortion or loss of information. This question seems to have been ignored in previous works about selective encryption using DCT like [Krikor…09], [Puech…05], or [Pareek…06].

In order to measure possible distortion or information loss in image processing, authors usually use Peak Signal-to-Noise Ratio (PSNR) [Huynh-Thu…08].

In our implementation, since DC coefficients are bounded by 2040, we used an 11-bit storage space. For AC coefficients, we have two sizes: the first one (called 11-bit storage in Figure 9(b) and Table 5) is using 11 bits with the first bit for the sign and the other 10 bits for the value (10-bit can store integers from 0 to 1023). The second size is using 8 bits (called 8-bit storage in Figure 9 (c) and Table 5) with a similar convention: the first bit for the sign, the other bits for the values (up to 102). The two storage methods are designed for different requirements of images quality.

If we take back one of the images of Figure 7 as our original image in Figure 9(a), we can observe the respective visual effects of using the 11-bit storage vs using an 8-bit storage in Figure 9(b) and Figure 9(c) respectively.

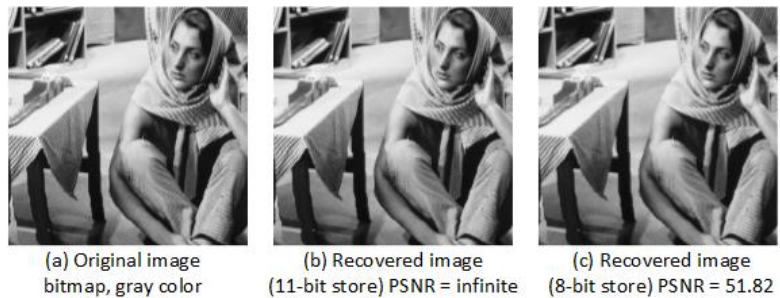

Figure 9 : PSNR and visual effect of different storage methods.

| Image size (in pixel) | 11-bit storage | 8-bit storage |
|---|---|---|
| 256×256 | Infinite | 51.74 |
| 512×512 | Infinite | 51.82 |
| 1024×768 | Infinite | 51.76 |
| 1600×1200 | Infinite | 51.72 |
| 3240×2592 | Infinite | 51.85 |
| 4800×4800 | Infinite | 51.87 |

Table 5 : PSNR for the selective encryption of different image sizes.

Table 5 provides the PSNR calculated for different image sizes for the 11-bit storage and the 8-bit storage. It is usually admitted that PSNRs of 50 result in almost identical images. When the PSNR is infinite, the two images are said identical [Huynh-Thu…08].

The extra storage space of 8-bit storage method is 5/64 +11/8/64= 9.96% (DC takes 11 bits and AC takes 8 bits) of the original image. However, extra storage space of 11-bit storage space is 11×6/64/8 = 12.9% of the original image. In this work, the 11-bit storage method is used as default; as pointed out in Table 5, this choice is making our DCT implementation virtually lossless for a wide array of image sizes. We will progress one more step (in sections 6.8 and 6.9) in this research of losslessness (aka integrity) by using DWT instead of DCT.

## 6.5. Implementation and Evaluation

In this section, we mainly discuss the implementation decision of allocating calculation tasks to GPU or to CPU. Then, we evaluate performance considering our two different hardware platforms: a laptop with limited computing capacity GPU and a desktop with a powerful GPU. Performances were so different that they led to different implementation decisions.

### 6.5.1. Allocation implementation decision regarding the laptop scenario

This implementation described in Figure 10 was already shown in our previous published work [Qiu…14]. The image data will be first copied into GPU memory where it will be fragmented after DCT 8×8 preprocessing. Then, the selected coefficients which are considered as the important part will be transferred to host memory and encrypted using AES 128-bit by CPU. In parallel, the rest of the DCT coefficients will be padded with zeros and transformed by IDCT 8×8 algorithm.

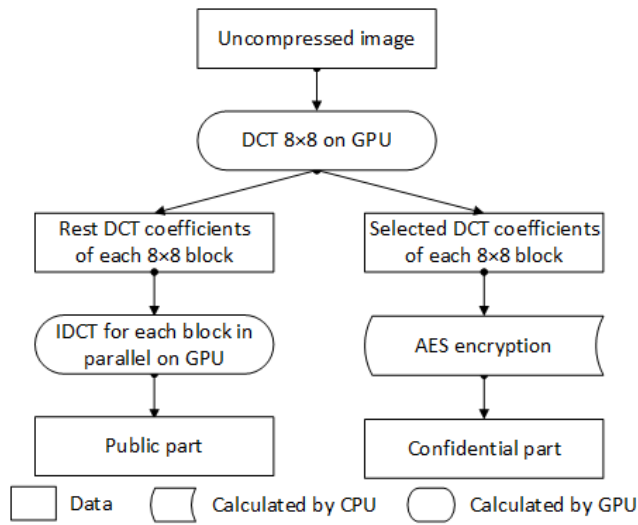

Figure 10 : Process steps for first level of protection.

The total run time depends on a race between CPU and GPU. The time spent using the GPU is greater than the time spent using the CPU as shown in the first rows of Table 6 and illustrated by Figure 11 where it can be observed that the CPU is idle most of the time.

This design based on the laptop hardware configuration works well for a series of images in the same format (bitmap) as input because of the parallel computation between the GPU and the CPU. As GPU are calculating the DCT 8×8 and IDCT 8×8 of each input image, the CPU is encrypting the selected confidential parts (data amounting to less than 10% of the original image) in parallel. The total run time depends on which processor is slower. The resulting flow of operations is shown in Figure 11 for two images I1 and I2.

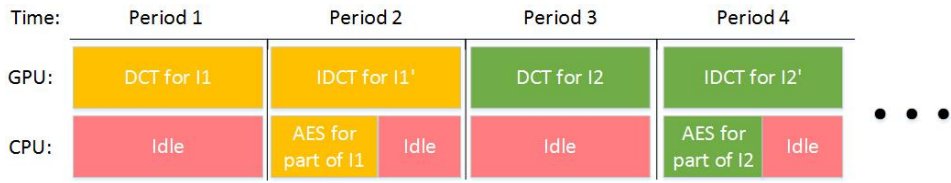

Figure 11 : Time overlay design for multiple bitmap images as series input.

In fact, as long as the encryption run time on CPU for part I1 is less than the total time of period 2 for I1 and period 3 for I2 as shown in Figure 11, this implementation decision of overlay will work perfectly which makes it well adapted for the strong level of protection implementation shown in Figure 12 as well. As shown in the last row of Table 6, the time consumed by SHA-512 /1024 algorithm on CPU can still be covered by the time consumed on GPU which makes this strong level of protection implementation have the same performance on a laptop as the first level of protection.

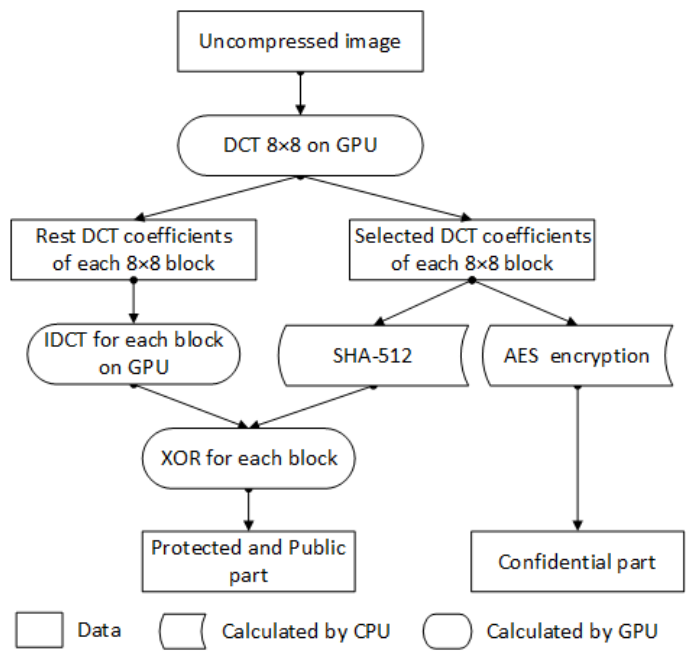

Figure 12 : Process steps for strong level of protection.

| Image size | 1024×768 | 1600×1200 | 3240×2592 | 4800×4800 |
|---|---|---|---|---|
| GPU time (DCT) | 0.41ms | 0.79ms | 3.67ms | 9.98ms |
| CPU time (AES) | 0.19ms | 0.47ms | 2.05ms | 5.87ms |
| CPU time (SHA-512) | 0.29ms | 0.73ms | 2.8ms | 7.69ms |

Table 6 : DCT time on GPU and AES time on CPU of the laptop.

The complete performance evaluation is shown in the third row of Table 7. The total runtime is not exactly twice the DCT runtime on GPU because DCT and IDCT are asymmetric and some coefficients are selected and padded with zeros. Our evaluation shows that the total runtime is over 1.1GB/s.

As shown by [Li…12] and [Gervasi…10], GPUs can accelerate AES computation. However, as shown in Table 2 and Table 3, GPU calculation capacity varies widely according to their architecture. In Table 7, performances of full image encryption using AES on CPU or on GPU are compared with our SE method. It is worth pointing out that due to limitation of the GPU computation capability on laptop, the AES on GPU is slower than on CPU.

| Image Size | 1024×768 | 1600×1200 | 3240×2592 | 4800×4800 |
|---|---|---|---|---|
| AES on GPU | 5.5ms | 13.5ms | 59.2ms | 162.3ms |
| AES on CPU | 2.1ms | 5.0ms | 21.9ms | 60.2ms |
| SE on CPU + GPU | 0.89ms | 1.94ms | 8.38ms | 20.9ms |

Table 7 : Speed of AES on GPU, on CPU, our SE design on CPU + GPU on a laptop.

### 6.5.2. Evaluations for first level protection on desktop

Table 8 and Figure 13 are showing that GPUs on some desktops are so powerful that they are able to calculate DCT (or IDCT) faster than the CPU is able to compute AES for the selected part of the image. The huge potential calculation resources of modern GPU makes it possible to do all SE steps including DCT 8×8 and AES on GPU. According to [Li…12], AES speed can reach more than 50 Gbps on an Nvidia GPU of a desktop machine with CUDA implementation. Our implementation was able to reach almost 40 Gbps on our desktop GPU as shown in Table 9. In such a situation, the parallel implementation of Figure 11 is not suitable anymore and we are rather getting a scheme like in Figure 13 instead where the GPU is getting idle time, the CPU is fully used during the period 2 and is on the critical path.

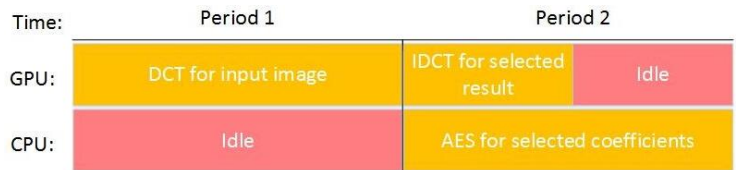

Figure 13 : Time overlay design on PC with a very powerful GPU.

Actually on desktop scenario, the CPU speed in time period 2 (AES for selected data) is much slower than the GPU speed (IDCT for rest DCT coefficients). In the desktop we used an Nvidia GeForce GTX 780 as a GPU and an Intel I7-4770K as a CPU (mentioned in previous section GPU acceleration). Table 8 shows performance for the period 2 of Figure 13.

| Image Size | 1024×768 | 1600×1200 | 3240×2592 | 4800×4800 |
|---|---|---|---|---|
| Period 2 on GPU | 0.04ms | 0.09ms | 0.41ms | 1.12ms |
| Period 2 on CPU | 0.16ms | 0.38ms | 1.72ms | 4.67ms |

Table 8 : Run time in period 2 on desktop GPU and CPU.

We then move all calculations of SE including DCT and AES to GPU. This design makes GPU to work three steps for each input image: DCT for original image, AES for selected part and IDCT for the rest coefficients. And we list the evaluation for full encryptions on CPU and GPU to be compared with SE on CPU and GPU in Table 9.

| Image Size | 1024×768 | 1600×1200 | 3240×2592 | 4800×4800 |
|---|---|---|---|---|
| AES on desktop GPU | 0.19ms | 0.46ms | 1.91ms | 5.46ms |
| AES on desktop CPU | 1.56ms | 3.8ms | 16.6ms | 45.5ms |
| SE on GPU | 0.10ms | 0.21ms | 1.01ms | 2.76ms |

Table 9 : Speed of AES on CPU and GPU, our SE (first level of protection) on GPU.

We can see that the SE we use is still faster than AES on either CPU or GPU. This improvement benefits from the design we discussed in last subsection that is move all the calculations of SE to GPU. Based on these implementations, we can see that using GPU as an accelerator for our SE algorithm is always a better choice compared with AES judged by the performance they have.

### 6.5.3. *Evaluations for strong level of protection on desktop*

Regarding the strong level of protection, we have to determine where to perform the hash calculation. According to tests based on programs described in [Steube 2013], SHA-512 performance on the desktop GPU, Nvidia GeForce GTX 780 is around 136 MH/s (meaning 136 million hash calculations per second). We should notice that for each 8×8 block, there will be one hash calculation, so we can estimate the run time by SHA-512 as in Table 10.

| Image Size | 1024×768 | 1600×1200 | 3240×2592 | 4800×4800 |
|---|---|---|---|---|
| Time consumed by SHA-512 once per 8×8 block | 0.09ms | 0.22ms | 0.96ms | 2.65ms |

Table 10 : Speed of SHA-512 of once per 8×8 block on desktop GPU.

The speed is much faster than SHA-512 CPU based implementation from (Dai 2009). Moreover, SHA-512 implementation on CPU is too slow to fit with the overlay design shown in Figure 13 in our desktop scenario. We conclude from this analysis that it will be more efficient to compute the hash calculation on GPU. At last, we provide with Table 11 a comparison between the strong level of protection and AES both developed on GPU.

| Image Size | 1024×768 | 1600×1200 | 3240×2592 | 4800×4800 |
|---|---|---|---|---|
| AES on desktop GPU | 0.19ms | 0.46ms | 1.91ms | 5.46ms |
| SE on GPU | 0.19ms | 0.43ms | 1.97ms | 5.41ms |

**Table 11 : Evaluation of AES on GPU, our SE (strong level of protection) on GPU.**

Table 11 shows that the strong level of protection on desktop with a powerful GPU available, has about the same performance as AES 128-bit for different images sizes. In summary, the allocation of calculation tasks of DCT, HASH and AES depends on the calculation capacity of the GPU. In fact, we can observe through tables 2 and 3 that GPU architectures progress at a faster pace than CPU architectures. Therefore, architectural decisions taken for the desktop implementation should prevail over time as powerful GPUs will eventually become more affordable and be present in laptop as well. The laptop architecture should remain interesting for energy critical environment though.

### 6.6. Protection evaluation by statistical analysis

In this section, histograms and correlation coefficients computation are used to evaluate the quality of protection relatively to the strong level of protection. As pointed out before, our method will fragment the image into two (a confidential fragment to be stored locally or in a high-level security place, a public fragment to be stored in a public server) and use the strong protection method as default to protect images.

It is well known that many ciphers have been successfully cryptanalyzed with the help of statistical analysis and several statistical attacks have been devised on them [Pareek…06]. To convince oneself of the robustness of our encryption method, two kinds of statistical analysis are performed by computing histograms then correlation coefficients for two adjacent pixels first in the original image of Figure 7(a) then in the corresponding encrypted public fragment of this original image.

#### 6.6.1. Histogram analysis

An image-histogram illustrates how pixels in an image are distributed by graphing the number of pixels intensity level. Figure 14 shows that the histograms of the cipher image are fairly uniform and significantly different from the respective histograms of the original image. This property guarantees that the cipher image will not provide any clue to be employed by any statistical cryptanalysis [Pareek…06].

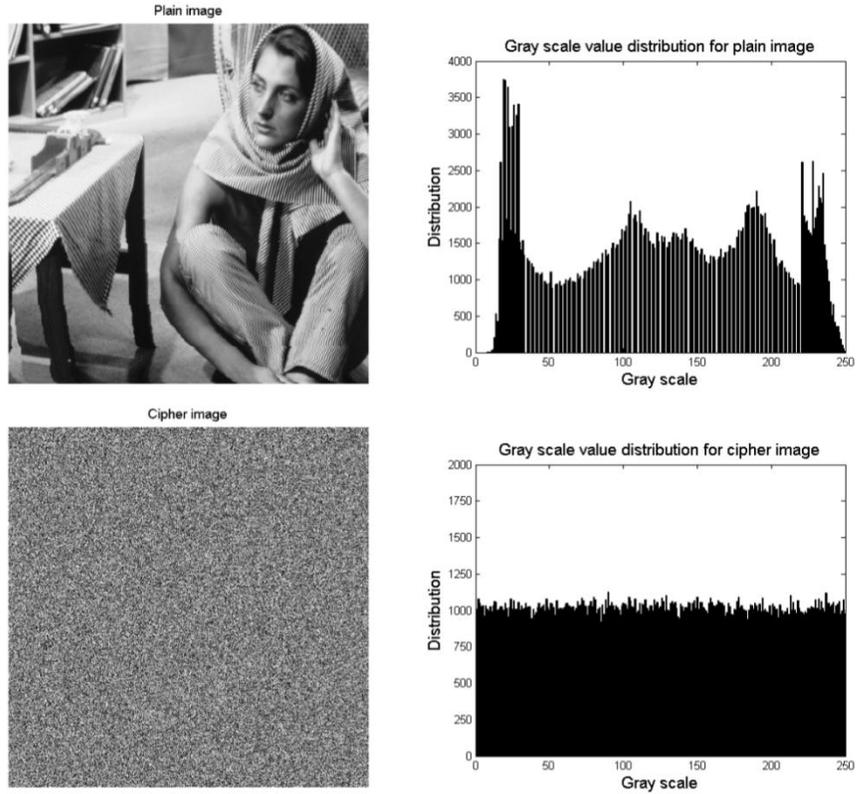

**Figure 14 : Plain image and its histogram compared with cipher image and its histogram.**

### 6.6.2. Correlation coefficient analysis

According to [Wang …11], to test the correlation between two adjacent pixels, the following procedures are carried out. First, randomly select 10,000 pairs of two horizontally adjacent pixels from an image and then compute the correlation coefficients $r_{xy}$ of each pair using the following classic equations:

$$cov(x,y) = E\{(x - E(x))(y - E(y))\},$$
$$r_{xy} = \frac{cov(x,y)}{\sqrt{D(x)}\sqrt{D(y)}},$$

where x and y are grey-level values of the two adjacent pixels in the image. Then, the same operations are performed along the vertical and the diagonal directions. As shown in Figure 15, the correlation coefficients of the cipher-images are very small.

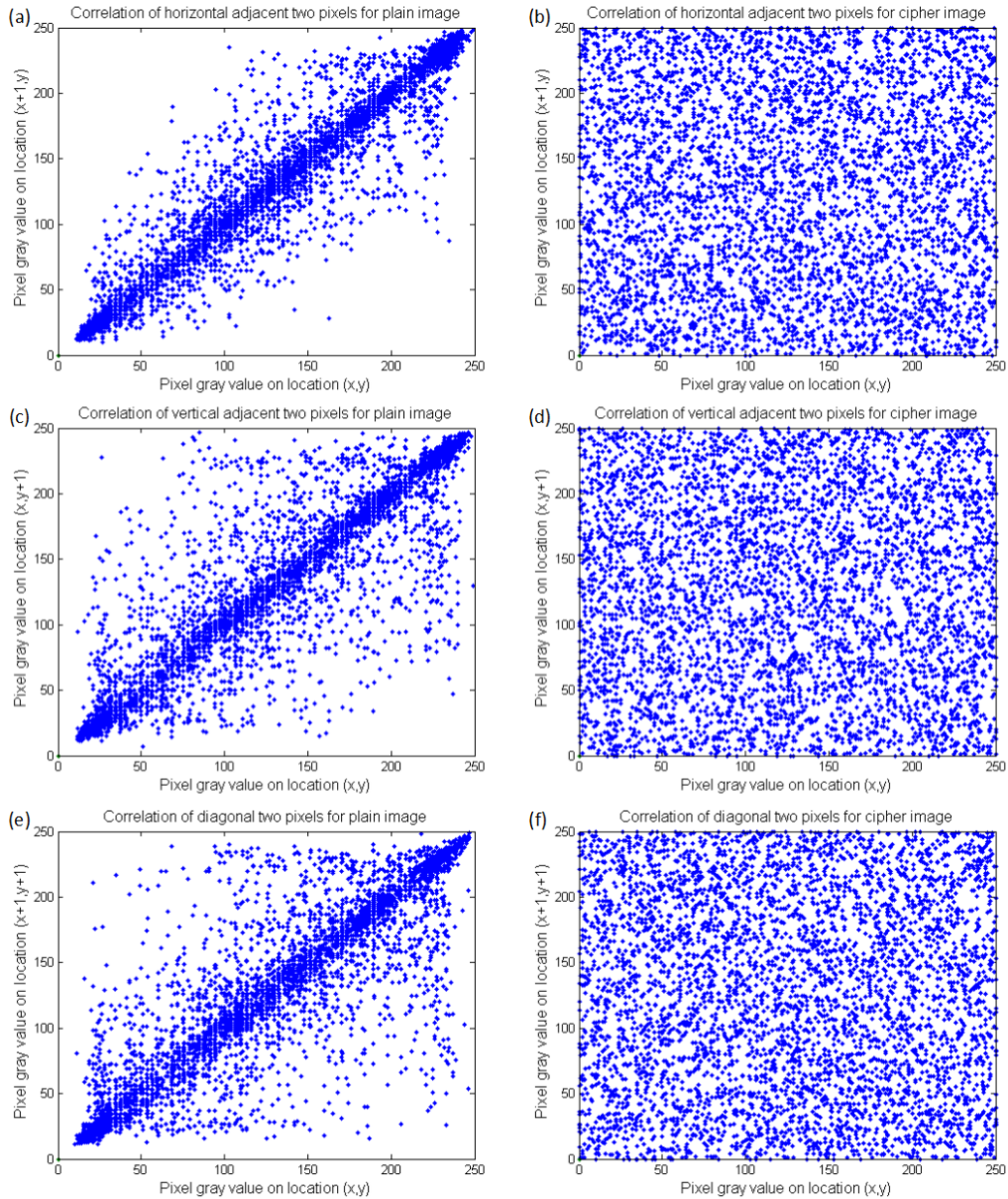

**Figure 15 : Correlation of adjacent pixels in horizontal, vertical and diagonal direction.**

### 6.7. Conclusion relative to selective encryption using DCT

In this work, we implemented two selective encryption methods both using DCT 8×8 preprocessing and GPU acceleration. We defined a first level of protection which is a continuation of our previous work [Qiu…14] and a novel strong level of protection. The first level of protection implementation combines calculation resources from both CPU and GPU available on most PCs, tablets, or even smartphones today. It provides a very fast speed to perform selective encryption in the frequency domain for uncompressed images like bitmap. We have seen on Table 7 and Table 9 along an array of different image sizes that it runs between 2 and 8 times faster than AES according to which GPU architecture is used. The strong level of protection addresses the issue with better protection for the public fragment left plain in the first level of protection. The idea is to use a small number of high frequencies to rapidly encode the low frequencies of the public fragment; indeed, the strong level of protection implementation also uses the acceleration offered by the GPU. Table 7 shows that it is still about twice faster than AES's on a laptop; and Table 11 shows that performances are comparable to AES's with a powerful GPU as the one found on a desktop. By two different statistical analyses, it has been possible to show that this method offers a good level of protection.

The separation of an image data into a public fragment and a confidential fragment can be used to address the issue with efficiently protecting large amount of bitmap images using but not completely trusting remote storage servers like a storage provider. We separate the original data as 'the' important confidential fragment to be stored locally and put the remaining data to a remote server for instance, in a cloud with the additional protection offered by the cloud provider. In doing so, we make the best usage of the local memory where we store only about 13 % of the image, depending on a tunable number of coefficients selected to constitute the confidential fragment. To perform one or the other of the two methods, we refined the implementation architecture using both the GPU and the CPU available on a PC and reach a level of performance up to about four times the performance of AES and never slower.

Indeed, one has to realize that GPUs architectures as well as encryption algorithms are progressing at a fast pace. For instance, late in 2014, a new generation Nvidia Geforce series GPUs (http://www.nvidia.com/object/geforce_family.html) was released with more CUDA cores, higher clock frequency and wider memory bandwidth, improving effective speed by 40% compared to GPU for the desktop we used. We are convinced that performance for computing the DCT 8×8 and other algorithms benefiting from GPU like SHA-512 or even AES will still progress. As pointed out by [Gregg…11], the memory transfer between host and GPU memory could be a bottleneck due to the limitation of the PCIE (Peripheral Component Interconnect Express) bus connecting them (normally several GB/s). During this work, we have seen that unfortunately, this can influence the load to assign to the CPU vs. the GPU in order to obtain the best performance. This would suggest developing a smart adaptor to allocate the computation task according to the hardware architecture available.

Nonetheless, our work is clearly showing that selective encryption can potentially become widely used for bitmap image protection since it now provides excellent processing time, a strong level of protection possibly fragmented in two separate storage space, a moderate increase of its total memory storage. In our opinion, selective encryption of an image must be considered as a step towards a more complex combination of encryption and fragmentation. We started to experiment with selective encryption for a bitmap image as described in [Massoudi…08a] or [Krikor…09].

### 6.8. Selective encryption using DWT

In the previous sections, DCT (Discrete Cosine Transform) is used to support fragmentation decision before performing encryption for image protection. As pointed out, although with a remarkable benchmark brought by the GPGPU, DCT cannot guarantee the losslessness due to conversions between integers and floating point numbers which will result in rounding errors. The rounding errors can be reduced by using more storage space but cannot be totally avoided. This problem makes DCT cannot provide the integrity required for dealing with other kinds of data.

Discrete Wavelet Transform (DWT) [Burrus…98] is used in selective encryption [Gonçalves…15], [Sadourny…03], [Pommer…03] but most of the time, it is rather used as a standard compression step for formatting rather than as a preprocessing step for selecting in multimedia use cases. In our design, the Le Gall 5/3 filter is used as it has an important lossless property by mapping integers to integers.

Benchmark or performance against full encryption is still needed. In some use cases, the preprocessing step of SE can legitimately be ignored as SE and compression are integrated and that transform is used by both applications. They just do light weight protection within the compression or coding process like in MPEG4 or JPEG2000. However, our use case encompassing any kind of data will have to take into account the entire process when it comes to performance evaluation. This will lead us to implement DWT on a GPGPU to benefit from the acceleration supplied by the parallel architecture of a GPU.

## 6.9. Design and key architectural choices

### 6.9.1. Discrete Wavelet Transform

DWT is a signal processing technique for extracting information mostly used in compression standard such as JPEG2000 or MPEG-4. It can represent data by a set of coarse grain and detail values in different scales. Naturally, it is a one-dimensional transform. But it also can be used as a two-dimensional transform as applied in the horizontal and vertical directions. For the image case, this will generate four small images which each one is one quarter the size of the original image with one level transform: one with low resolution (LL), one with high vertical resolution and low horizontal resolution (HL), one with low vertical resolution and high horizontal resolution (LH), and one with all high resolution (HH). Then, the second level transform will only be done for the first quarter of the first level's result which is called dyadic decomposition as shown in Figure 16.

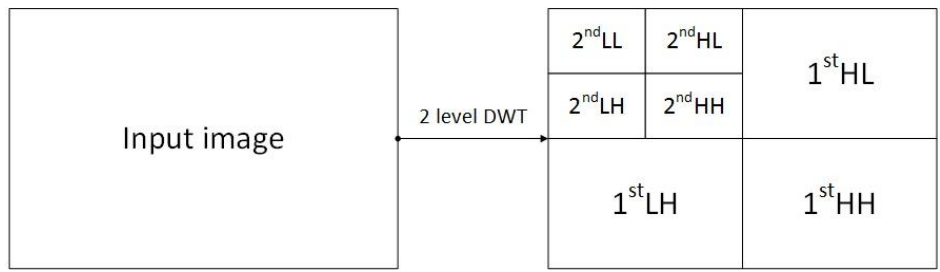

Figure 16 2-Level DWT generates two-dimensional coarse and detail values [Burrus...98].

To perform the forward DWT, a one-dimensional subband is decomposed into a set of low-pass samples and a set of high-pass samples. By using LeGall 5/3 filter [Burrus...98], no data will be altered due to numerical rounding. And the lifting-based filtering is used which updates odd sample values with a weighted sum of even sample values, and updating even sample with a weighted sum of odd sample values. The lifting-based filtering for the 5/3 analysis filter is achieved by using (1) and (2) below:

$$y(2n+1) = x_{ext}(2n+1) - \lfloor \frac{x_{ext}(2n) + x_{ext}(2n+2)}{2} \rfloor \quad (1)$$

$$y(2n) = x_{ext}(2n) - \lfloor \frac{x_{ext}(2n-1) + x_{ext}(2n+1) + 2}{4} \rfloor \quad (2)$$

where $x_{ext}$ is the extended input signal, y is the output signal and $\lfloor a \rfloor$ indicates the largest integer not exceeding a.

We chose a two-level DWT as illustrated in Figure 16. The selected coefficients to build the private fragment are the $2^{nd}$ LL which takes about 1/16 of the storage space and carries the basic elements (coarse information) of the original image. The reason of using two-level DWT is that the one-level DWT still has a large part (1/4 of the storage space) of LL coefficients to protect and any other level DWT makes the selected coefficients very few such that the remaining public fragment could unveils too much information.

### 6.9.2. GPGPU

In order to improve DWT performance and make good use of all the capabilities of a PC's hardware, the General Purpose Graphic Processing Unit (GPGPU) with a CUDA (Compute Unified Device Architecture [Kirk 07]) support is utilized and greatly accelerates the computation of DWT. A CUDA algorithm performs the one level 2D DWT (2D Daubechies 9/7) in 45ms (without data transfer between GPU and host memory) on a 4096x4096 image using an Nvidia Tesla C870 which is about 20 times faster than a PC CPU of the same hardware generation.

As a GPU is designed to parallelize the calculation tasks on a hardware level [Kirk 07], the DWT-2D on tiled $8 \times 8$ blocks fits better on a GPU than a CPU. In order to fully use the calculation resource of a PC, the CPU is also used for calculation depends on the allocation of different hardware platform. In our implementations, we used the same two different GPUs used with the SE based on DCT with characteristics described in Table 3.

| GPU Type | Computation capacity version | CUDA cores | Memory (MB) | Clock (MHz) | Memory Width |
|---|---|---|---|---|---|
| Nvs 5200M | 2.1 | 96 | 1024 | 1344 | 64 bit |
| GeForce gtx 780 | 3.5 | 2304 | 3072 | 941 | 384 bit |

Table 12 Main characteristics of a laptop GPU and a desktop GPU

Unlike traditional CPUs which have only several powerful physical cores (normally 4 or 8 on a Intel CPU for PC) that allows only limited number of threads in parallel physically, a GPGPU could contain hundreds even thousands of threads execution physically in the same time. The most commonly used example is the CUDA architecture from Nvidia [Kirk 07]. However, the GPGPU calculation capacity has a phenomenal growth in recently years. The Nvidia Nvs 5200M GPU is a low-end GPU on a laptop with only 96 CUDA cores and 1GB GPU memory while the GeFroce gtx 780 GPU is a high-end GPU released in 2014 with 2304 CUDA cores and 3GB GPU memory. The calculation capacity between the two GPUs have a huge difference which the GeForce gtx 780 is about 10 times faster on Hash (evaluated by [Steube 13]) and DWT calculation tasks than Nvs 5200M shown in Table 13.

| GPU Type | SHA-256 (MH/s) | SHA-512 (MH/s) | DWT-2D 8 × 8 (GB/s) |
|---|---|---|---|
| Nvs 5200M | 49.9 | 11.3 | 0.95 |
| GeForce gtx 780 | 585.1 | 101.2 | 6.97 |

Table 13 Hash and DWT

### 6.9.3. Design of SE method for arbitrary nature of data

In our design, as shown in Figure 17, in order to deal with sizable input data it is being proposed to cut into several chunks of the same given size 2D matrix (e.g. seen as a 512 × 512 or 1024 × 1024 pixels gray color image which is chosen to accommodate further transformation or the hardware platform architecture). Then every chunk ($D_i$) goes to the SE process to generate two fragments: the private fragments $D_iA$ and the public ones $D_iB$. Then the $D_iA$ parts go to a local machine under the user's control and the $D_iB$ parts may be transmitted to the public area like a public cloud with no fear of an attack.

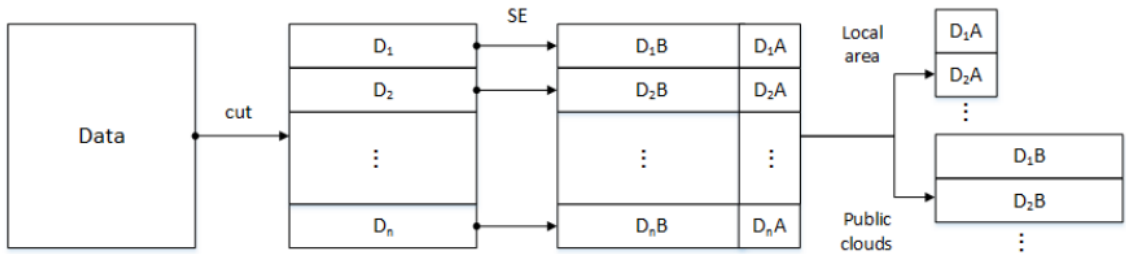

Figure 17 SE General method of processing large amount of data.

The main idea is to consider every chunk $D_i$ as a matrix and be treated as such by the SE process. That is to say any kind of data can be seen as a matrix by considering every byte of data as a pixel to form a bitmap gray scale image. Then every chunk $D_i$ is simply processed using the SE method block by block with block size 8x8 as shown in Figure 18. The block size chosen can be changed according to the size of the original data. This tiling step is used for fitting with the GPGPU architecture (which will be mentioned later in this paper).

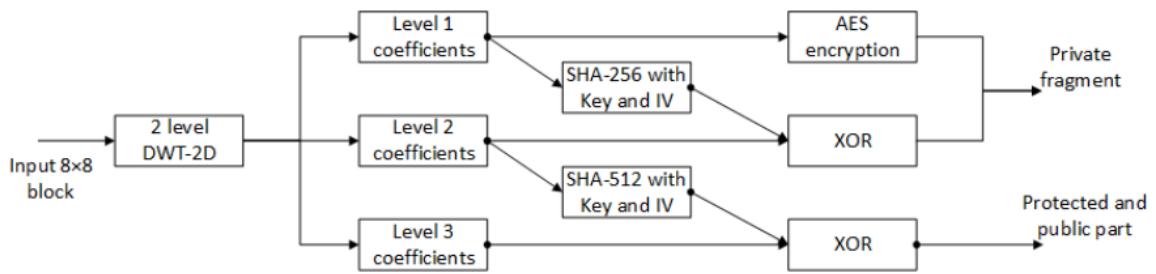

Figure 18 SE for the single 8x8 block of one image.

The first step for the 8×8 block is to perform the Discrete Wavelet Transform (DWT). In our work, two successive levels of the DWT are performed so the low frequency coefficients which are considered as the most important part (level 1 coefficients in Figure 18) and will be protected by AES-128 bit take only k out of 64 coefficients (with k = 4 in our implementation) but carry most of the original frequency feature. Then the level 1 coefficients will generate a bit sequence with a length of 256 to protect the level 2 coefficients (the rest coefficients of 2nd level DWT) by performing a XOR operation also with the AES key and IV (Initial Vector). The SHA-256 is used here to generate very different 256-bit sequence even when the selected coefficients of the neighbor blocks are very similar. These two parts of coefficients are considered as private and should be stored in a trusted area (locally in our scenario). In our design, however, the code is structured such that another cipher algorithm can easily replace AES-128 if need be. Of course, if this machine was to be considered as a trusted area then any kind of protection for the selected coefficients including no protection would do. Then the level 2 coefficients will be used to generate a bit sequence with a length 512 using SHA-512 also with AES key and IV to protect the level 3 coefficients (rest DWT coefficients) by performing a XOR operation also with AES key and IV. As long as the low frequency coefficients of DWT represent the basic elements of an image, the protected most part of the high frequency coefficients can be stored on public clouds without any information leak.

### 6.9.4. Numerical precision and storage space usage

The preprocessing step used to separate data for the SE, the data before and after the preprocessing step could be very different. This could lead to the difference of the storage space usage or rounding errors caused by conversions between integers and floating point numbers. In [Guan…05], the authors claim all variables are declared as type *double* with a bit-length of 64 bits. This is unfair if the input data are stored as integers especially *int* type with a bit-length of 8 bits as the storage of the results will require 8 times more storage space compared with original data. In [Qiu…15], we described how to optimize integer representation but still could not avoid possible rounding errors caused by the calculation of DCT.

In this section, the preprocessing step is the DWT based on LeGall 5/3 filter which is designed to be an integer-to-integer map, such that the DWT is lossless. As a result, on one hand, any rounding error can be avoided; on the other hand, the extra storage space usage caused by the int to float conversion does not exist. The only possible extra storage usage could be caused by the different value range of the input 8-bit int and the output int coefficients. And the output value range can be calculated as long as the input values are always stored Byte by Byte, the input value range (seen as unsigned value) is from 0 to 255 which can be considered as from -128 to +127 (the range is seen as from -128 to +128 during the following calculation). Then the storage methods can be designed according to the value range distribution.

The first level DWT-2D transform is actually calculated by twice DWT-1D transforms (equation (1) and (2)) on the 8x8 block in horizontal and vertical directions sequentially. The first horizontal transform generates two sub-matrices which are $1^{st}$L and $1^{st}$H that take each half of the result matrix horizontally. The vertical transform is done on each of the two sub-matrices which generates four sub-matrices like in Fig. 4 ($1^{st}$LL, $1^{st}$HL, $1^{st}$LH, $1^{st}$HH).

In the first horizontal transform, the range for 1stH is -255 to +255 (double the range of input) and the range for the 1stL is -192 to +192 (1.5 times of the input range). Then the transform in vertical direction, which is transform of the 1stL and 1stH blocks respectively, gets the following results: 1stHH is from −511 to +511 and the range for 1stLH is from −384 to +384.

All the coefficients in the three sub-matrices of first level DWT- 2D transform can be stored using 10-bits storage space.

The value range of second level DWT-2D coefficients are generated by the same two direction DWT-1D transform of the $1^{st}$LL submatrices coefficients. Range of the second level DWT coefficients can be estimated by simplifying the equations (1) and (2) and then directly get results from calculating final formula of each elements in the four submatrices in Figure 18 ($2^{nd}$LL, $2^{nd}$HL, $2^{nd}$LH, $2^{nd}$HH). And the max values and min values for each of the value estimated are shown in the following matrices. And the storage method for the second level DWT-2D coefficients is: 11-bits for each of the lower left corner four coefficients and 10-bits for rest of the coefficients.

$$\begin{bmatrix} 338 & 267 & 468 & 468 \\ 267 & 211 & 369 & 369 \\ 468 & 369 & 648 & 648 \\ 468 & 369 & 648 & 648 \end{bmatrix}, \begin{bmatrix} -338 & -267 & -468 & -468 \\ -267 & -211 & -369 & -369 \\ -468 & -369 & -648 & -648 \\ -468 & -369 & -648 & -648 \end{bmatrix}$$

As pointed out in Figure 18, the confidential fragment we selected are two levels of coefficients which is the four submatrices ($2^{nd}$LL, $2^{nd}$HL, $2^{nd}$LH, $2^{nd}$HH) and the $2^{nd}$LL is the level 1 coefficients. The storage environments for the three levels of coefficients could be flexible. If the level 1 and 2 are kept local and only level 3 coefficients are stored in clouds, the avalanche effect [Kirk 07] can be avoided (This is the design we chose to evaluated in this paper). However, if the channel is reliable and transmission mistakes are rarely to see, the level 2 coefficients can also be put on clouds without leaking any information about the level 3 coefficients so the storage locally is very little. The confidential fragment will take 164-bits storage in total (40-bits for level 1 and 124-bits for level 2) and the protected and public fragment takes 480-bits. The total storage space usage is 644-bits at least which is about 26% more than original but the most part can be stored on clouds without leaks and the important part is at most 32% according to the stronger storage design.

In summary, the preprocessing step is the DWT-2D based on LeGall 5/3 filter which is designed to be an integer-to-integer map, such that the DWT is lossless. As a result, on one hand, any rounding error can be avoided; on the other hand, the extra storage space usage caused by the int to float conversion does not exist. Moreover, in our design, we consider any kind of data type as int with bit-length of 8 bits. That is to say, no matter what kind of original data type it is, we process the data by reading byte seat a time and deal with it as an 8-bit integer. Then the input bytes will form an "image" (2-D matrix) of a configurable size ready for the whole SE process. In this process, the numerical data type of variables involved in DWT computation is carefully designed such that it can provide integrity for any kind of input data.

### 6.10. Security Analysis

A secure encryption algorithm ought to resist existing powerful attacks [Nyberg…95], [Cho…11]. In this section, different security tests on the proposed scheme are performed to establish its high level of security. Indeed, as the protected data is divided into two parts and each one are protected. The first private part, which has a small size, is encrypted by AES-128 and stored locally in our scenario, while the second public one is stored on the cloud while permits to reduce the stored overhead locally and provide protection for the data stored in clouds.

The basic assumption is the selected private part of data is secure by using AES-128 (also, it is easy to replace AES-128 with any other encryption algorithms as in Figure 18), so the security property of this part is not analyzed in this section. To validate the safe employment (robustness) of the proposed method, the public part which can be stored on clouds in our use case is analyzed in terms of cryptographic performance to verify if it reaches the required cryptographic performance.

In the following, we present some of the figures for the security analysis but all statistical results can be found in Table 14. Moreover, as long as both image and text file are used as test examples, some criteria are just suit for images while not text files; some criteria are not applied for text files as input.

### 6.10.1. Statistical Analysis

Immunity against statistical attack requires that the cipher must reach a high level of randomness. To validate the robustness of this cipher, different statistical security tests are applied such as uniformity of encrypted image and independence between plain and encrypted images.

| Statistical results for images | | | | |
|---|---|---|---|---|
| | Min | Mean | Max | Std |
| PSNR | 9.1946 | 9.2464 | 9.3062 | 0.0173 |
| SSIM | 0.0068 | 0.0101 | 0.0133 | 0.0010 |
| $Dif$ | 49.7923 | 49.9995 | 50.1790 | 0.0620 |
| $KS$ | 49.7945 | 49.9971 | 50.1859 | 0.0627 |
| $\chi^2$ | 190.2125 | 256.0844 | 333.9687 | 22.3298 |
| $H-O$ | 4.2769 | 4.2769 | 4.2769 | 0.0000 |
| $H-E$ | 5.7581 | 5.7656 | 5.7722 | 0.0022 |
| $\rho 2$ | -0.0108 | -0.0001 | 0.0108 | 0.0036 |
| $\rho-h$ | -0.0560 | -0.0002 | 0.0603 | 0.0168 |
| $\rho-v$ | -0.0593 | 0.0002 | 0.0621 | 0.0165 |
| $\rho-d$ | -0.0550 | 0.0002 | 0.0603 | 0.0161 |

| Statistical results for texts | | | | |
|---|---|---|---|---|
| | Min | Mean | Max | Std |
| $Dif$ | 49.3762 | 50.0016 | 50.6113 | 0.1819 |
| $KS$ | 49.3937 | 50.0058 | 50.5652 | 0.1837 |
| $\chi^2$ | 185.466 | 254.5267 | 325.0422 | 23.0735 |
| $H-O$ | 4.5191 | 4.5191 | 4.5191 | 0.0001 |
| $H-E$ | 5.758 | 5.7657 | 5.774 | 0.003 |
| $\rho$ | -0.0291 | 0.0001 | 0.0304 | 0.0101 |
| $NMI$ | 0.0678 | 0.0623 | 0.0714 | 0.0006 |

Table 14 : Statistical results of sensitivity for interesting part (stored locally) for Lenna image (a) and a random text (b)

### 6.10.2. Uniformity Analysis

The encrypted data should possess certain random properties such as the uniformity, which is essential to resist against frequency attacks. Therefore, the PDF (Probability Density Function) of the encrypted data should be uniform. This means that each symbol has an occurrence probability close to 1/n, where n is the number of symbols (1 256 = 0.039 in byte level). We start by analyzing the image data and then text data to prove that the proposed method can attain the uniformity independently for its public part.

The original plain image Lenna and its corresponding PDF are shown in Figure 19-(a), (b). While, in Figure 19-(c), (d), the corresponding cipher image that is stored in cloud (c) to their corresponding PDF (d) is shown, respectively. It can be observed that the PDF of the encrypted images using the proposed scheme is close to uniform distribution since the probability of different symbols is close to 0.039.

Also, in Figure 20, the byte representation of an original chosen text file is presented in (a) and its corresponding PDF in (b) in addition to its corresponding encrypted text byte representation that is stored in cloud (c) and with its corresponding PDF (d). The result indicates that the encrypted text file also possesses a uniform distribution. From these results, we have shown that the distribution of encrypted data tends to the uniform one no matter of the input data type. Moreover, to validate this result, an entropy test is realized in the sub-matrix level of size 8×8 (same size as the plain-text input block).

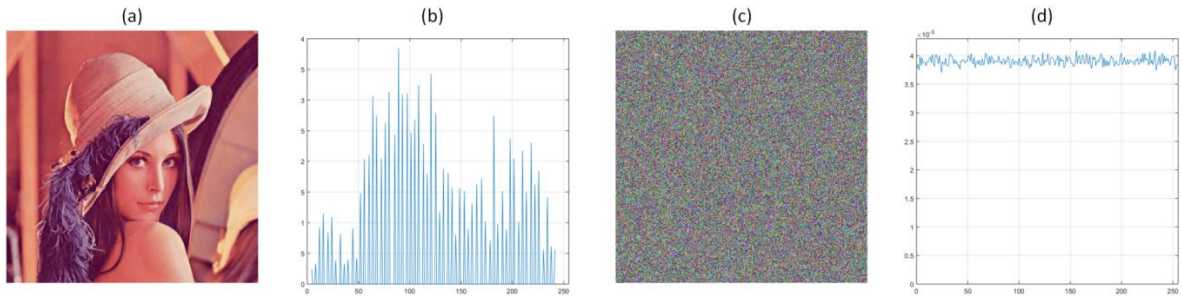

**Figure 19 (a) Original image, (b) PDF of the original image, (c) encrypted image, (d) PDF of the encrypted image**

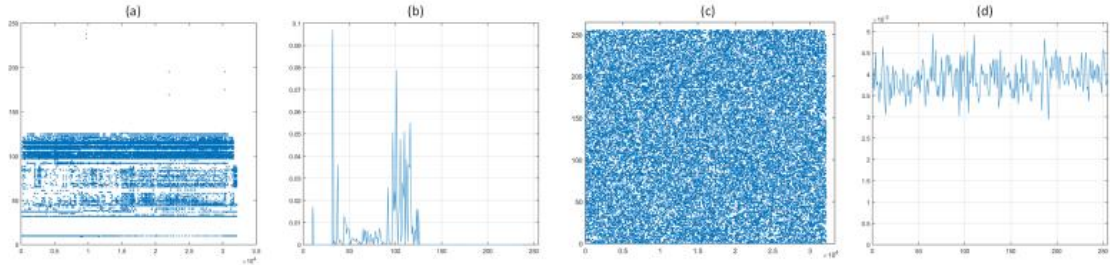

**Figure 20 (a) Original text byte representation, (b) its corresponding PDF, (c) Encrypted text stored in cloud, (d) its PDF corresponding representation**

### *6.10.3.   Information Entropy Analysis*

The information entropy of a data sequence M is a parameter that measures the level of uncertainty in a random variable and it is expressed in bits, and is defined using the following equation (3):

$$H(m) = - \sum_{i=1}^{n} p(m_i) \log_2 \frac{1}{p(m_i)} \qquad (3)$$

Where p(mi) represents the occurrence probability of the symbol mi and n is the total states of the information source. To the best of our knowledge, the related works applied the entropy test in the level of image. However, this is not significant. We propose to apply the entropy test in the level of sub-matrix with size h × h (same size of input and encrypted plain-text block). Indeed, each sub-matrix can be considered a truly random source with uniform distribution if it has an entropy equal or close to $\log_2(h^2)$. The value of entropy close $\log_2(h^2)$ is the desired value.

$$H(m) = -\sum_{i=1}^{h^2} \frac{1}{n} \log_2 \frac{1}{h^2} = \log_2(h^2) \qquad (4)$$

The entropy analysis for the original and for the encrypted Lenna stored in cloud is shown in Figure 21-(a) for h = 8. These results indicate that the encrypted sub-matrices always have an entropy close to the desired value, which is 6 in case of h = 8. Also, the same test is applied for the text file shown in Figure 20-(a) and the result is shown in Figure 21-(b).

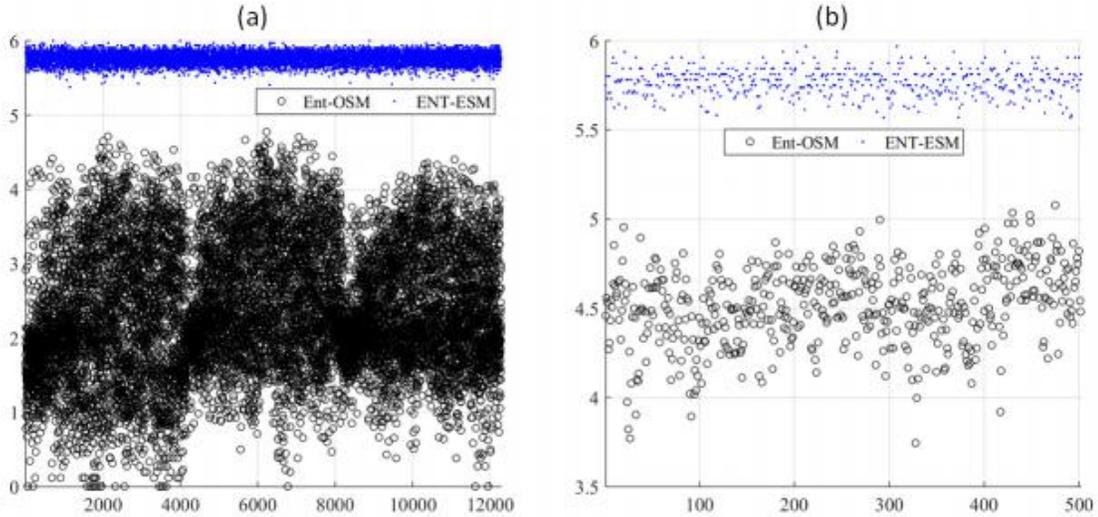

Figure 21 Entropy analysis (for h= 8) for the sub-matrices of (a) original and encrypted images , of (b) text

Moreover, the average entropy for the encrypted Lenna image versus 1000 random secret keys is 5.766 and for the text file is 5.766 which are both very close to 6 (see H − E in Table 14) while are also very different from the original entropy (see H − O in Table 14). Therefore, the proposed cipher is secure against entropy attack in both image and text cases.

### 6.10.4. Correlation test between the Original Data and the Public Fragment

A low level of correlation between original and encrypted data is an important factor that allows validating their respective independence. Having a correlation coefficient close to zero means that the high degree of randomness is reached. The correlation coefficient $r_{xy}$ is calculated using the same method as in section 6.6.2.

In this test, we also use image and text files as input for analyzing potential correlation. Indeed, the variation of coefficient correlation between original and encrypted matrices is obtained by applying 2-D correlation coefficient and the result is shown in Table 14 (see value distribution of $\rho^2$ for image case and $\rho$ for text case). The result indicates that the 2-D coefficient correlation varies in a small interval around 0. This means that low 2D-correlation coefficient is reached by employing the proposed cipher approach and consequently the independence between the original and encrypted matrices is statistically attained.

Additionally, to validate that spacial redundancy is removed for the encrypted image, the correlation between pixels of original and encrypted images are realized. This test selects randomly N = 4096 pairs of two adjacent pixels in horizontal, vertical and diagonal direction. The results are presented in

Figure 22, for the original (a)-(c) and encrypted Lenna image (d)-(f) in horizontal, vertical and diagonal directions, respectively (same for text file from (h) to (m)). The result in this figure clearly indicates the high correlation between adjacent pixels in plain matrix input (correlation coefficient close to 1). While, for the ciphered matrix, the correlation coefficients become very low (close to 0) which clearly shows that the proposed scheme reduces severely the spatial redundancy.

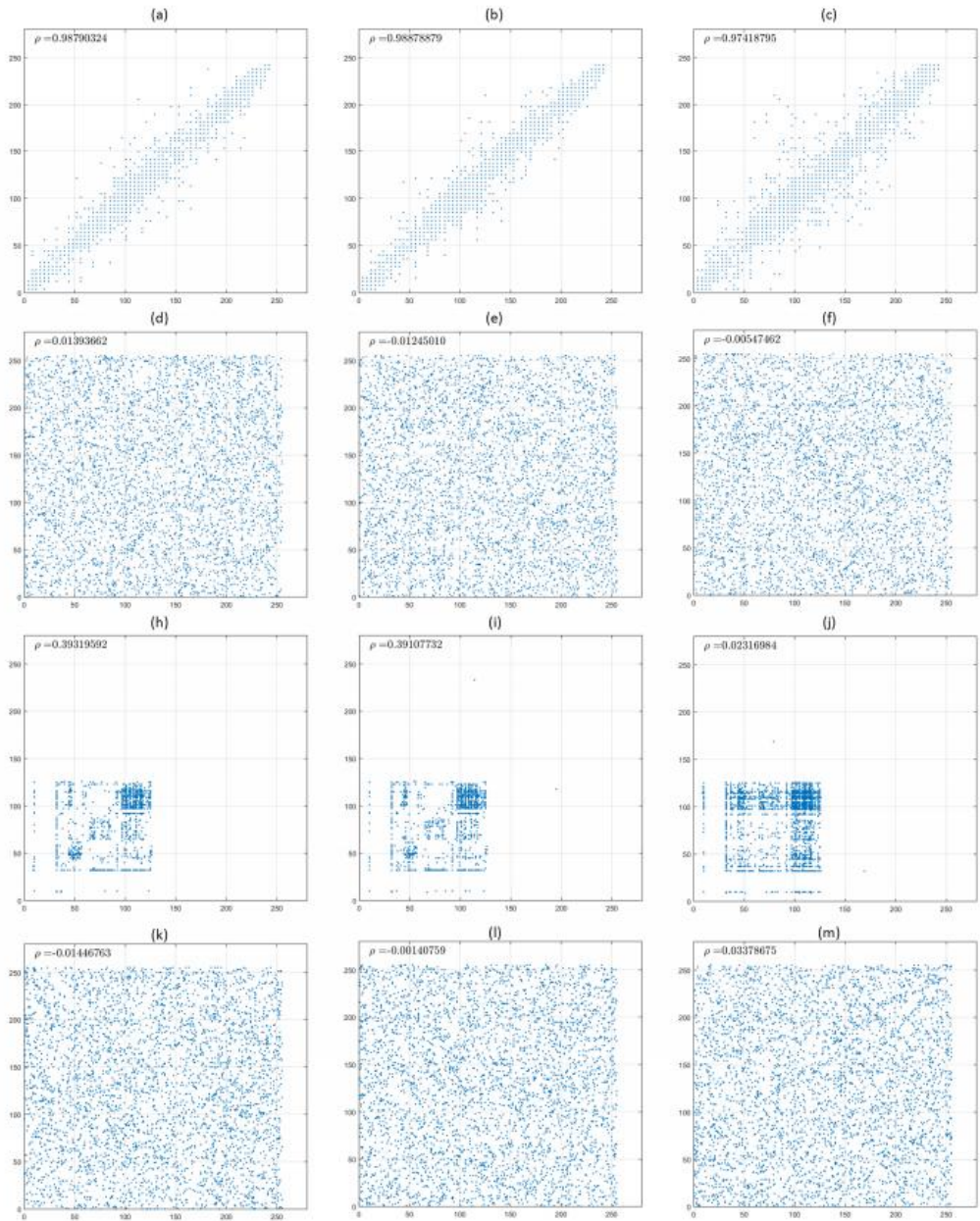

**Figure 22** *Correlation distribution in adjacent pixels in the original image (a) horizontally, (b) vertically, (c) diagonally Correlation distribution in adjacent pixels in the encrypted image (d) horizontally, (e) vertically, (f) diagonally Correlation distribution in adjacent pixels in the original text (h) horizontally, (i) vertically, (j) diagonally Correlation distribution in adjacent pixels in the encrypted text (k) horizontally, (l) vertically, (m) diagonally*

Moreover, the variation of the correlation coefficient between adjacent pixels of encrypted Lenna image versus 1000 random keys are shown in Table 14 ($\rho-h$, $\rho-d$, $\rho-v$ respectively. The results are close to 0, which confirms that spatial redundancy is eliminated and no detectable relation can be found in the encrypted matrix for both image and text case. Similar results are obtained using text file as input (see

Figure 22 (h)-(m)).

### 6.10.5. Difference Between Plain and Ciphered Images

The encrypted data must be as different as possible from the original data, at least 50% in a bit level comparison. The proposed scheme achieved a high value of difference before and after processing for any data format. For example, the plain image Lenna was tested and the result in Figure 23-(a) shows that 50% of bits is being changed between the encrypted and the plain image. Additionally, similar result is obtained for text file and statistical value is shown in Table 14 (see value distribution of Dif for image case and text case).

Additionally, in order to confirm this result, we applied the Normalized Mutual Information (NMI) between the original and encrypted matrices of 1000 random secret keys and the results (in Figure 23-(b) for an image, in Table 14 for a text) are showing that NMI is always close to 0 (mean equal to 0.0193). Consequently, this strongly indicates that no meaningful information can be extracted from the encrypted data.

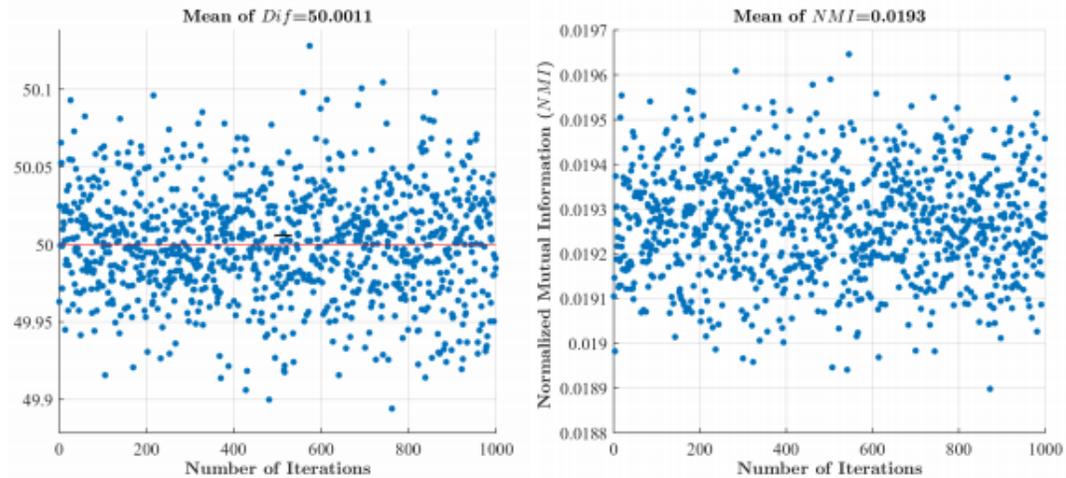

Figure 23 Difference (a) and MNI (b) between original image and encrypted image using 1000 random different keys

### 6.10.6. Sensitivity Test

Differential attacks are based on studying the relation between two encrypted data resulting from a slight change, usually one different bit in the original plain image or in the key. A successful sensitivity test shows how much a slight change will affect the cipher data. In other words, the higher the ciphered data changes when slight change happens in input, the better sensitivity of the encryption algorithm is. Here, we analyze different types of sensitivity.

For the Plain-text Sensitivity, it is designed that different Initial Vector (IV) is employed for each input matrix which leads to produce totally different cipher image for the same plain image. Therefore, the cipher has successfully met the avalanche effect but in a different manner and based on the key and IV sensitivity.

Concerning the Key and Initial Vector Sensitivity tests, it is one of the most important tests and allows quantifying its sensitivity against any slight change in the secret key or in the initial vector. In fact, the proposed key derivation function is based on the mixed of secret key and an initial vector, which means that similar sensitivity performance should be reached. To study the key sensitivity, two secret keys are used: S K1 and S K2 that differ in only one random bit. The two plain-images are encrypted separately and the Hamming distance of the corresponding encrypted Lenna images C1 and C2 is computed and also for the chosen text file, and illustrated as Table 14 (see KS for both cases) versus 1000 tests.

It is seen that the obtained values are always close to the optimal value (50 %) for the both input data. This indicates that the proposed method ensures high sensitivity against any tiny change in the secret key. Similar results are obtained for the IV sensitivity.

### 6.10.7. Visual Degradation

This test is specific for image and video contents and allows quantifying the visual degradation reached by employing a cipher scheme. In fact, the degradation operated on the original image must be done in a way that the visual content presented in the ciphered image must not be recognized. Two parameters are usually studied to measure the encryption visual quality which are Peak Signal-to-Noise Ratio (PSNR) [Hyunh…08] and Structural Similarity index (SSIM).

PSNR is derived from the Mean Squared Error (MSE), while MSE represents the cumulative squared error between the original and encrypted images. A lower PSNR value indicates that there is a high difference between the original and the cipher images.

The SSIM index is defined after the Human Visual System (HVS) has evolved so that we can extract the structural information from the scene. SSIM is in the interval [0,1]. A value of 0 means that there is no correlation between the original and the cipher images, while a value close to 1 means that the two images are approximately the same. In this context, PSNR and SSIM are measured between the original and the encrypted Lenna images for 1000 different keys and corresponding value distribution presented in value did not exceed 0.036, which means that a high and hard visual distortion is obtained using the proposed cipher algorithm.

Figure 24, respectively. The mean PSNR value is 9.23 dB which validates that the proposed cipher provides a high difference between the original and the encrypted image. Also, the SSIM

As a conclusion, the proposed cipher scheme ensures a hard visual degradation. This means that no useful visual information or structure about the original image could be revealed from the cipher image.

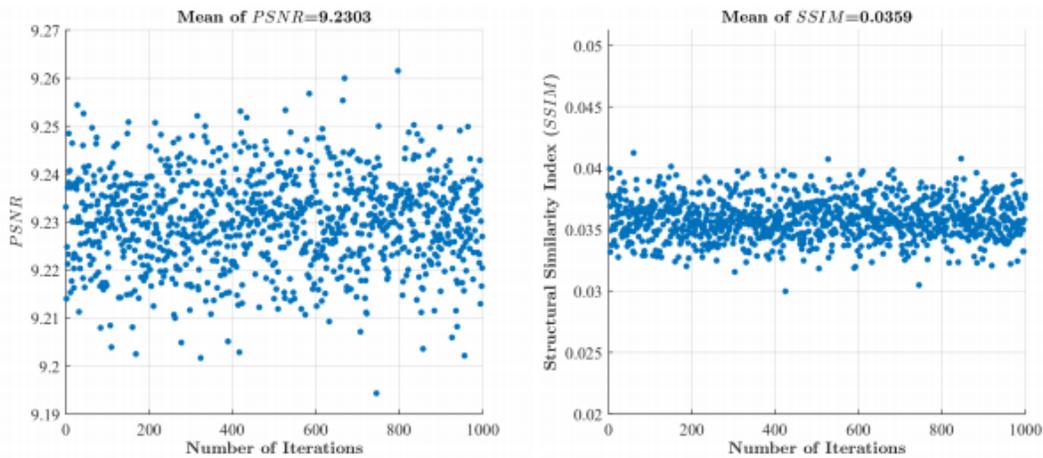

**Figure 24 PSNR and SSIM variation between the original and the encrypted image using 1000 random different keys**

### 6.10.8. Error propagation

An important criteria that should be ensured for any cipher is the error propagation. The interference and noise in the transmission channel (or in the store system) might cause errors. Bit error means that a substitution of '0' bit into '1' bit or vice versa. This error may propagate and lead to the destruction of decrypting data, which is a big challenge since a trade-off between avalanche effect and error propagation are shown in [Massoudi…08a]. In this proposal, if a bit error takes place in the encrypted cipher sub-matrix, the error will propagate randomly only in its corresponding sub matrix and will not affect its consecutive corresponding decrypted sub-matrix. Moreover, as we presented before, in Figure 18, the Level 3 coefficients are the only part that transmits. This means the decrypting error could only happen when XOR the Hash results of Level 2 coefficients and protected Level 3 coefficients. The decrypting data will have 1 bit error if there is 1 bit error in the protected Level 3 coefficients during transmission. As a result, we can conclude that the proposed approach is efficient to overcome error propagation.

### 6.10.9. Cryptanalysis Discussion: Resistance against the well-known types of attack

In this part, typical cryptanalytic cases appearing in the literature are considered and a brief analysis of the proposed cipher against several cryptanalytic attacks is provided from a cryptanalysis viewpoint. The proposed method is considered to be public and the cryptanalyst has complete knowledge to all steps but indeed no knowledge about the secret key and the dynamic initial vector (Kerkhoff's principle).

The strengthen of the proposed scheme against attacks is based on the dynamic key that produced in function of a secret key and a dynamic secret initial vector by applying a hash function (one way property) for each input plain-text. Indeed, it is very difficult for an attacker to recover the dynamic secret key that is changed for every input data. Therefore, the problem of single plaintext failure and accidental key disclosure is avoided by this scheme. Furthermore, differential and linear attacks would become ineffective. In fact, any change in any bit of the secret key or IV (public parameters) causes a significant difference in the produced encrypted image as seen in Table 14. Moreover, we apply a robust block cipher AES with a secret key length of 128, 192, or 256 bits, which is sufficiently large to make a brute-force attack utterly difficult. Hence, a sufficient secret key is used and as the difficulty of cipher-text-only attack is equal to one of the brute force attacks, it becomes impossible for a cipher-text-only attack to retrieve useful information from the public part in our scheme. Therefore, our method resists the cipher-text attack.

Toward resist the statistical attacks, the proposed approach achieves that the plain-text are changed in positions and values, which means that the confusion and diffusion properties are ensured in addition. An example is illustrated in Figure 25, where a 8×8 matrix of the original Lena image and its corresponding cipher one is illustrated by their values. This result demonstrates that all values are changed. Therefore, the randomness property is ensured and this consequently permits to prevent the reverse-attack algorithm.

More importantly, the spatial redundancy between adjacent elements of input plain data are removed and a high randomness degree of the whole encrypted data are proved. Different statistical tests such as the entropy analysis, probability density function, correlation tests are applied to validate the independence and uniformity proprieties. Consequently, these results indicate that no useful information can be detected from the public part. This validates the robustness of the proposed approach and their high resistance to statistical attacks.

Moreover, key sensitivity analysis demonstrates the efficiency of the proposed cipher scheme against related key attacks, while any change in any one bit of key provide a different (50%) encrypted data.

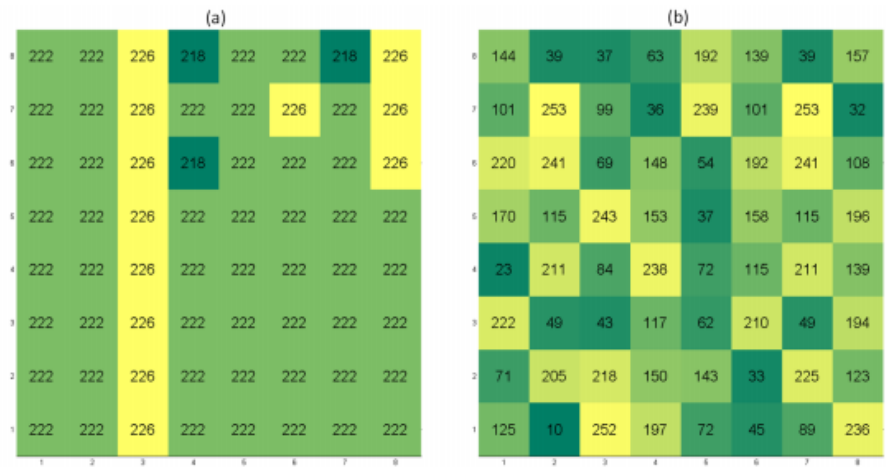

Figure 25 (a) 8x8 cropped plain image with its corresponding gray scale matrix, (b) its encrypted version

## 6.11. Performance evaluations

In this section, we evaluate performance of the whole protection process. As we are considering the allocation of the calculations on a PC platform, the hardware resource we have are a CPU and a GPGPU. However, the calculation capacity of GPGPU could be very different that changes the whole execution time of SE [Owens…08]. So the performance is evaluated in two typical use cases that are a laptop equipped with a low-end GPU and a desktop equipped with a high-end GPU.

### 6.11.1. Evaluation for all calculation tasks

A key implementation decision is to distribute the calculation tasks between the GPU and the CPU. As pointed out in Section 2.3.2, the DWT-2D, SHA-256 and SHA-512 can benefit from the GPU acceleration, so the design is based on the parallel execution of CPU with GPU while the GPU will take calculation tasks of DWT- 2D, SHA-256 and SHA-512 in the process while CPU takes AES-128 for only Level 1 coefficients. The initial plan for both low-end and high-end GPU cases is to keep the GPU busy and CPU would have idle time space available for other tasks (Figure 26).

In the laptop, there are an Intel I7-3630QM CPU and an Nvidia Nvs 5200M GPU. For the desktop, we have an Intel I7-4770K CPU and an Nvidia Geforce gtx 780 GPU. In order to verify our initial plan and allocate the right calculation tasks to the right chip, we evaluate each of the tasks on laptop and desktop and results are shown in Table 15. For different size of input data chunk, the execution time of GPU for the second input data chunk can always overlap the execution time of CPU for the selected DWT-2D coefficients of the first input data chunk.

From the two hardware environments, we evaluated, the overlay design in Figure 26 works. And the speed of the whole SE process relies on how fast the GPU can process its calculation tasks on input data as long as there are many chunks as input. That is to say, in these two scenarios, the time consumed by GPU is evaluated as the benchmark for our SE method. The calculation speed evaluated for this laptop scenario is about 360 MB/s and for this desktop scenario is about 2.8-3.2GB/s.

|  | Image Size | 1024 × 1024 | 2048 × 2048 | 3200 × 3200 | 4800 × 4800 |
|---|---|---|---|---|---|
| Laptop Scenario | GPU time (DWT-2D) | 1.39ms | 4.87ms | 12.6ms | 24.1ms |
| | GPU time (SHA-256) | 0.33ms | 1.31ms | 3.3ms | 7.3ms |
| | GPU time (SHA-512) | 1.45ms | 5.8ms | 14.2ms | 31.8ms |
| | CPU time (AES-128) | 0.29ms | 1.14ms | 3.06ms | 6.67ms |
| desktop Scenario | GPU time (DWT-2D) | 0.34ms | 0.79ms | 1.7ms | 3.3ms |
| | GPU time (SHA-256) | 0.05ms | 0.13ms | 0.29ms | 0.63ms |
| | GPU time (SHA-512) | 0.13ms | 0.69ms | 1.58ms | 3.2ms |
| | CPU time (AES-128) | 0.23ms | 0.96ms | 2.37ms | 5.3ms |

Table 15 Performance evaluation for every calculation taks of SE for two different Hardware environments for different image sizes

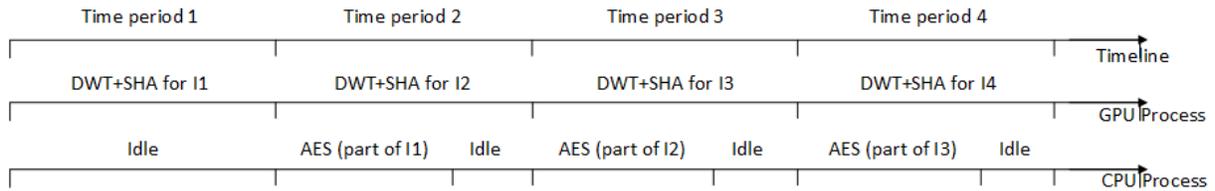

Figure 26 Load distribution between GPU and CPU processors

### 6.11.2. Discussion on performance

We used two very different GPU architectures with contrasted characteristics (see Table 12): 2304 CUDA cores for the desktop GPU to be compared with the 96 CUDA cores on the laptop GPU. These differences are sufficient to explain the differences in performance between the laptop and the desktop. As long as GPU architectures are rapidly evolving, the hardware configuration of a GPU may strongly influence software applications architectural choices necessary to derive the best possible code. This point could even invert the results of our evaluation since a large number of cores could very well favor the GPU calculation tasks that finally change the overlay design. Also as pointed by [Gregg…11], when a GPU calculation capacity is very high, bottleneck of the process is the memory transfer between the GPU memory and the host memory instead of calculation speed itself.

Here we presented the comparison of the SE method with the traditional CPU-only AES-128 speed ([Dai 09]) in Figure 27. It is worth noticing that [Bogdanov…15] pointed out the CPU-only AES could also be very fast with the support of the New Instructions extension brought by Intel. This AES-NI could accelerate the AES on CPU for more than 5 times and achieve almost 3GB/s on a desktop CPU (shown in Figure 27). In short, as long as the GPU is evolving rapidly, the computation design could always change to achieve the best performance.

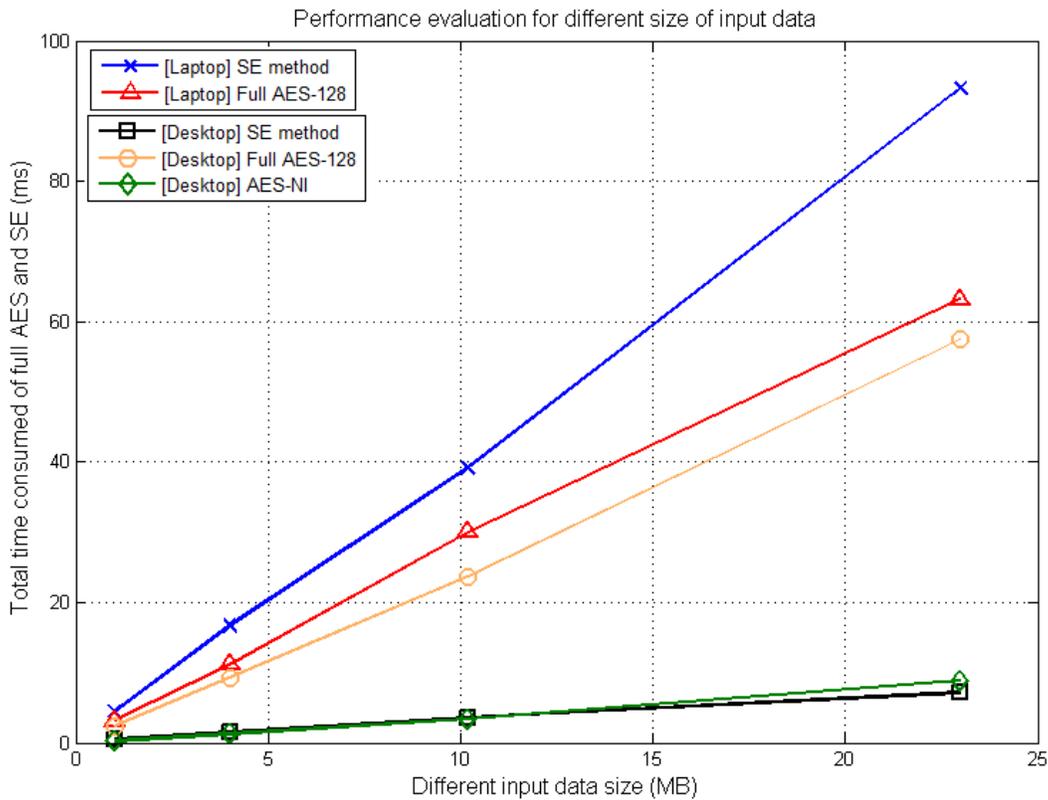

**Figure 27 Time overlapping architecture of the implementation**

### 6.12.   Conclusion relative to selective encryption using DWT

In this paper, an efficient agnostic selective encryption algorithm was presented. It separates the input data into two fragments. A first fragment called private, contains most of the information within about 25% of the memory size of the original data. The separation preprocessing must be fast, guarantee integrity, and enable memory optimization at the same time. This is attained by employing the invertible DWT-2D $8 \times 8$ preprocessing implemented by using a GPU and that allows   accurately recovering the plain-text without any error.

The proposed scheme can be applied to any kind of data and not only to multimedia contents at the difference of any SE method that we have been able to see in the current literature. The proposed approach employs the robust block cipher AES with counter mode to protect the interesting part that will be stored locally. In addition, the plaintext of the interesting part is hashed with a counter to produce a key-stream matrix. This matrix will then be employed to encrypt the non-interesting part by mixing it with the produced key-stream. The encrypted non-interesting part will be transmitted to the cloud and according to the obtained results, no useful information can be detected, which ensures that the data confidentiality and the privacy of users. Moreover, the proposed scheme can be realized in parallel to permit the parallelism processing and consequently reduce latency. More important, we recommend employing GPU to reduce the overhead of applying the optimization DWT-2D operation. To validate that the proposed design can ensure the required goals, a benchmarking was realized between the proposed solution and a full CPU on two different kinds of data. Several experimental and theoretical analysis were realized to prove the efficiency, robustness and resistance against error propagation. Therefore, the proposed solution can be considered as a good selective encryption algorithm candidate that can be applied for any system and for any kind of data.

# 7. Non selective fragmentation of data into several fragments

Our analysis now focuses on textual data stored in distributed storage solutions exploiting data fragmentation and dispersion as a way of providing additional security. Today, data fragmentation is widely used for resilience and scalability purposes: for instance, it can be found in the RAID technology [Raid], as well as in clusters like Apache Hadoop [Hadoop]. However, the fact of dividing data into pieces and dispersing those pieces on multiple machines in order to constitute a considerable obstacle for an attacker is not yet widely spread.

Fragmentation for security can be found in a theoretical paper from late 70's [Shamir 79] where it addresses the problem of secure storage and management of an encryption key. Then, a more architectural description can be found from the 80's [Fray…86] with a design breaking sensitive data from non-confidential fragments and dispersing them on separate physical devices. During following decades, fragmentation idea was applied to relational database systems [Aggarwal…05], [Ciriani…10] and more recently, fragmentation was used in the context of the cloud technology [Bkakria…13], [Hudic…12], [Vimercati…13]. Today, we are facing new challenges along scalability and exposure: Petabytes of data are to be protected over large distributed systems of thousands of machines, which can at least partially, be public.

In this report, we propose to organize existing storage systems using fragmentation for security purpose into two groups. The first group addresses the need for archiving data structures on storage architecture without making any assumption about the type of data or about the kind of storage servers (they are all considered identical in term of trustworthiness). This group includes both academic and commercial solutions. In the second group, confidential data are stored on secured devices in contrary to non-sensitive data that can stay in public areas. We analyze both categories of systems in terms not only of security, but of memory occupation, resilience, key management, and performance (including latency). We conclude this section by discussing some fragmentation issues and recommendations

## 7.1. Bitwise data fragmentation and dispersion

This section presents an overview of systems using data fragmentation and dispersion without any consideration for their structure or nature, their semantic, or their uneven level of confidentiality. They just consider data as a sequence of bits, this is why we qualified them of '*bitwise*'.

Table 16 and Table 17 summarize their most important features with regard to data dispersion: secrecy, key management, availability, integrity and recovery in case of loss of information, defragmentation processing, number of fragments sufficient for an attacker to recover data, and required storage space. The first three systems are based on secret splitting techniques and guarantee information-theoretic security: an attacker will not be able to find out any information about a single fragment until he obtains the corresponding one-time pad for that fragment. Two commercials solutions, Cleversafe and Symform, are only computationally secure (it is possible to obtain some information about a fragment while having enough of time and computational resources), but significantly optimize the storage space by combining AES encryption [AES 01] with data dispersion. Most of these systems are dedicated to archiving or long-term storage. They are efficient for larger rather than smaller data objects.

### 7.1.1. Encryption techniques used for dispersed storage

This subsection presents the most common ways of dividing data into fragments, which do not reveal any information until being regrouped.

Secret sharing schemes, also known as threshold schemes, fragment data of size d into n pieces of the same size d of the original data, in such a way that any k pieces are sufficient to recover the original information while k-1 are not. In contrary to standard encryption techniques, secret sharing schemes provide information-theoretic security and do not require key management. One of the most popular sharing schemes is the Shamir's threshold scheme [Shamir 79] based on polynomial interpolation. Another scheme is the [Blakley 79] scheme relying on the fact that any n nonparallel (n-1)-dimensional hyperplanes intersect at a specific point. An XOR splitting is also a common and non-costly way of implementing secret sharing: to produce n fragments of the secret, n-1 random fragments are generated and one additional fragment is calculated, which XOR-ed with other fragments will give the secret back. As XOR splitting is a (n, n)-threshold scheme, it is not possible to recover the secret in case of loss of any fragment. However, because of their expensive memory consumption, secret sharing schemes are mainly used for protection of small sized data like encryption keys. In [Krawczyk 93], Krawczyk describes a storage efficient way of protecting information using a combination of the Shamir's secret splitting and the Rabin's dispersal algorithm [Rabin 89]. Rabin's scheme breaks data of length L into n pieces, each of length L/m, so that every m pieces suffice for reconstruction of the whole data. It is usually used for fault-tolerant storage and information transmission.

In Krawczyk's method, pieces of data are encrypted with a randomly generated key. Then, data is divided into n fragments using Rabin's algorithm and Shamir's threshold scheme is applied only for key protection purpose. A significant storage space saving was also achieved in [Parakh...10] – to reduce the size of fragments they applied redundancy to the Shamir's algorithm. However, with gain in storage space both algorithms, Krawczyk's, and Parakh and Kak's, lost the information-theoretic security property: an attacker disposing of enough time and huge computational resources can deduce some information from fewer fragments than required for data reconstruction.

### 7.1.2. Key management and storage space

Key management is one of the features presented in

Table 16 and Table 17. More fragments can imply more keys to be stored and managed. In systems that leverage secret splitting as a way of providing secrecy one plaintext fragment corresponds to two or more encrypted fragments. As a consequence, storage space in such systems is at least twice as big as the initial data size. Key management in this situation consists in assembling the right pairs or groups of fragments. Cleversafe and Symform base their protection on the AES encryption performed on fragments of data. The first commercial system comes with a so-called keyless solution, where the key is integrated in the data itself in an additional fragment. There is no need for key management: since once will obtain a sufficient number of fragments, he will be able to decrypt all of them. In Symform solution the key for each fragment is generated as a hash of the fragment itself. Keys are stored and managed inside a central trusted part of the cloud.

### 7.1.3. Data recovery

The possibility of data reconstruction in case of loss of a part of the information extends the longevity of a storage system. All of the outlined solutions from

Table 16 and Table 17 add some redundancy to countermeasure availability problems. Redundancy can be assured by a simple use of duplication of fragments, like in the case of POTSHARDS [Storer…09] and GridSharing [Subbiah…05], or by application of error correcting codes, which is the case of commercial solutions. The first way is more costly in terms of storage, but does not require complex calculation. Error correcting allows minimizing space on servers, but can be a burden for performance, especially in case of smaller data.

Another aspect that must be considered is the longevity of data while losing information about fragments location and order. Only POTSHARDS system addresses directly this problem by adding to fragments approximate pointers, which are references containing the possible location of other fragments of the same data. The architecture of remaining systems would allow a data recovery implying the cooperation of all of the storage nodes.

## 7.2. Academic and Commercial System Descriptions

### 7.2.1. PASIS

In 2000, the PASIS [Wylie…00] project adopted secret sharing for data protection in its implementation of a survivable storage system capable to handle storage nodes failures and malicious user activities.

PASIS uses p-m-n threshold schemes to encode data before it is stored. Data is broken into n fragments such that any m of them allows data recovery and fewer than p reveal no information. For maximum efficiency, the choice of an appropriate threshold scheme is adapted to the specific data to be protected, especially to its size. Moreover, an over-requesting technique speeds up the performance of data retrieval. To overcome the problem of malicious clients, PASIS comes with a self-securing storage: storage nodes internally version all data and audit all requests, thus providing time to detect intrusions. Repair agents installed on storage nodes are in charge of rebuilding damaged data based on the change history.

In comparison to a conventional primary-backup system, PASIS achieves better confidentiality, availability, durability and integrity. Its weak points are an excessive amount of required storage, as well as a significantly higher latency.

### 7.2.2. GridSharing

Another implementation of a distributed storage system comes from 2005 [Subbiah…05]. GridSharing combines perfect sharing scheme with replication of secret's fragments on different devices in order to build a fault-tolerant and secure distributed system fulfilling confidentiality, integrity and availability requirements. According to the authors, choosing perfect sharing schemes for data protection evades problems connected with key management.

An interesting idea of fragments renewal for additional protection is raised. The renewal consists on a periodical replacement of existing data fragments with new ones. This way the attacker has a shorter time interval for collecting all of the fragments required for data recovery. Perfect sharing schemes (where all the fragments are required to reconstruct the data, like the XOR splitting method) allow the implementation of efficient share renewal procedures: changing stored fragments without previous data defragmentation. For imperfect sharing schemes, where the knowledge of fewer than the threshold numbers of fragments can provide some information, a renewal algorithm has not been developed to date.

**Table 16**

*Comparison of bitwise data dispersion for academic storage systems. Blocks for compromise - number of fragments needed for data recovery, b -number of Byzantine faulty servers, d - initial data size, k - minimum number of fragments required for data reconstruction, l -number of leakage faulty servers, n - total number of fragments, n1 -number of fragments in the first split, n2 -number of fragments in the second split, N - number of servers, p - minimum number of fragments that reveals any information about data, r -number of servers in a row in the logical grid.*

|  | **PASIS** | **POTSHARDS** | **GridSharing** |
|---|---|---|---|
| Data encryption, secrecy | Various threshold schemes adapted to information type | Perfect sharing scheme: XOR splitting $n_1$ of $n_1$ | Perfect sharing scheme: XOR splitting |
| Data resilience, availability | Threshold schemes $p$-$k$-$n$ | Threshold splitting k of $n_2$ | Fragments replication |
| Key management | Depending on the chosen protection method | Keyless | Keyless |
| Integrity and recovery | Repair agents, auditing | Algebraic signature | Voting system |
| Defragmentation | Directory service maps objects and fragments. Fragment name contains the node location and the local name. | User knows shards indexes and data decomposition. In case of indexes loss: use of approximate pointers. | Voting system: user asks multiple servers for one fragment and chooses the most appropriate fragment from received answers. |
| Trusted element | PASIS agent integrated within the client system | Data transformation component | Client system |
| Blocks for compromise | k of n | k of $n_2$ for each of $n_1$ fragments | n |
| Storage space | nd | $n_1 n_2 d$ | $d \frac{r}{l+b} (r-1-b) \frac{N}{r}$ |

GridSharing takes into account the possibility that storage servers can leak information (they can reveal their content and state to an adversary, but execute the specified protocols faithfully) or be Byzantine (they can deviate from the specified protocol, they can also reveal their fragments) faulty, as well as the fact that some of them can crash.

It introduces a new l-b-c scheme, in which a l number of servers is leakage-only faulty, not more than b servers can be Byzantine faulty, and not more than c servers can crash.

The work shows high computation overhead of the Shamir's secret sharing scheme. A combination of two mechanisms is proposed in order to overcome inevitable performance problems. The first one implies the use of an XOR perfect sharing scheme, where all the fragments are needed for secret recovery. Replicating fragments on servers and implementing a voting system (for each share at least (2b+1) responses must be received and the value returned by at least (b+1) servers is the correct one) to determine incoherent fragments is the second mechanism.

Two allocation schemes of fragments on n servers are presented. In the direct approach servers are arranged in a logical grid having (l+b+1) rows, with at least (3b+c+1) servers in each row. Data is split in the same number of fragments as the number of rows. Then, each fragment is replicated along one row. In the second approach named GridSharing, each of the servers contains a couple of fragments. n servers are arranged in the form of a logical rectangular grid with r rows and n/r columns. As in the first approach, servers in the same row replicate the same data.

GridSharing system comes with stronger security than encryption-based techniques and an easy way of data sharing in collaborative environment. Dimensions of the architectural framework can be varied to trade-off between the number of required servers, the storage blow-up and recovery computation time.

### 7.2.3. Potshards

In 2007, the POTSHARDS [Storer…09] project from UCSC addressed the need of providing a secure archive that will be able to last for years. Its basic concept is to distribute the data between several cooperating organizations forming an archive system.

To assure archive longevity, the authors decided to use secret splitting schemes instead of encryption for two reasons. First of all, key management over years can be expensive and there is no guarantee that a couple of keys will not be lost. Secondly, even the strongest encryption is only computationally secure and can become easily breakable with the development of new technologies and within the long period of time as the objective is to preserve data for decades.

Before being stored in POTSHARDS archive, the data is processed in two splitting layers. The first splitting layer uses an XOR-based algorithm to generate random fragments from user data. Each of produced fragments reveals no information about the original data. In order to recover information, one needs to possess all fragments. Availability is the goal of the second splitting layer, which takes as input fragments coming from the first split. It applies the Shamir's threshold scheme and replaces each fragment with a group of shards - data pieces of the same length than the initial fragment. Only few shards are needed for the fragment reconstruction. Finally, shards are randomly distributed across independent organizations. POTSHARDS assures data integrity by the use of algebraic signatures.

Objects, fragments, and shards can be identified by their IDs. After data distribution, a user obtains a list of indexes corresponding to his archived objects. Because it is possible for this list to be lost, shards include additional portions of information called *approximate pointers*. Pointers of one shard show the archive region where shards from the same object are located. As a consequence, a user can recover data from shards even if all other information, such as the index, is lost. An intruder would have to steal all of the shards that approximate pointers refer to. This implies, among other things, to be able to bypass the authentication mechanisms of each archive.

In case of loss of a part of the stored data, a recovery is possible with the collaboration of all of the implied archives. While rebuilding data, participating archives reveal no information about their contents. First, archives agree on the destination of the data to be recovered by choosing a new fail-over archive. Then recovery process occurs in multiple rounds during which each of collaborating archive generates a random block and XORs it with a block of data needed for reconstruction. All the blocks generated in one round are XOR-ed together and send to the fail-over archive. At the end, the fail-over archive receives the random blocks used during the recovery process and calculates the lost data.

POTSHARDS system provides an information-theoretical security and does not require any key management. With enough time, it is possible to recover data even if shards location maps have been lost. Collaboration between organizations allows rebuilding a lost archive, but also implies the existence of a trade-off between secrecy and reliability. The cost of having a long-term archive is the amount of required storage: both the secrecy and availability splits are space consuming.

**Table 17**

*Comparison of bitwise data dispersion for commercial storage systems. Blocks for compromise - number of fragments needed for data recovery, d - initial data size, k - minimum number of fragments required for data reconstruction, n - total number of fragments.*

| | Cleversafe | Symform | SecureParser |
|---|---|---|---|
| Data encryption, secrecy (2.1) | Information dispersal: AONT transform with AES-256 | Information dispersal and AES-256 | Information dispersal and AES-256 |
| Key management | Key inside the data | Key stored in the Symform Cloud Control central element | Key inside the data |
| Data resilience, availability | Reed-Solomon error correction $k$ of $n$ | Reed-Solomon error correction 64 of 96 | Threshold splitting $k$ of $n$ |
| Integr | SHA-256 | SHA-256 | Authentication value inside each fragment |
| Defragmentation | Maps in Accesser switches indicate the location of fragments and the optimal storage node | Map stored in the Symform Cloud Control. | Map stored inside the Parsed File System. |
| Trusted element Blocks for compromise | $k$ out of $n$ | 64 out of 96 | $k$ out of $n$ including all mandatory shares |
| Storage space | $n\,d/k$ | $(1.5)d$ | $n\,d/k$ |

Cleversafe [Cleversafe] is one of the first commercial solutions implementing data dispersal as a way to provide security. Its dispersed storage network offers a complete software and hardware architecture for private cloud storage. Petabyte scalability, reliability and decreasing storage space are the key drivers of the project.

The data processing is done at the source or in a dedicated hardware element. While fragmenting, a user chooses the level of redundancy that he wishes to obtain. Then, data is transformed and encrypted using the AONT-RS approach [Rivest 97], which combines the All Or Nothing Transform [Resch…11] with Reed-Solomon erasure code [Reed…60]. First, the data is fragmented into words of the same length and each word is encrypted with a random key using the AES-256 algorithm. Subsequently, a SHA-256 hash value (a canary) of the data is generated in order to provide an integrity check. The next step is to create the last word by XOR-ing the canary with the key. It is not possible to reconstruct the data unless someone obtains all the fragments and retrieves the key from the last word. For availability purposes, Cleversafe applies a modified version of Rabin Information Dispersal Algorithm [Rabin 89] on the encrypted data, based on the Reed-Solomon erasure code. As a consequence, additional pieces of data are generated, so the user can reconstruct the data even if some of the pieces are lost. After the fragmentation, the data is dispersed on random storage nodes.

For security reasons, data must be secure not only at rest, but also in motion: if the attacker intercept the traffic and catches all the fragments, he can be able to reconstruct the data. To attain the objective of secure transport of fragments, Cleversafe verifies all nodes that would like to join its storage network. Moreover, transported data is protected by the use of encryption. No need for key management (because the key is included in the data) makes data management less costly. On the other hand, the key is transmitted and stored with the data. Once the attacker gains the access to the storage (for example by breaking the authentication mechanisms) or arrives to observe the fragments passing through the network, he will have all the elements needed for reconstruction of the original data. In Cleversafe, data protection relies in inability of the attacker to collect the data from multiple locations or to intercept the traffic of the fragments from the storage servers to the client.

The major drawback of the AONT-RS approach is its performance on small objects. Cleversafe introduces a separate way of processing for such data. The new approach abandons the use of time-consuming error correction code processing. Instead, data duplication is applied.

### 7.2.3. *Symform*

Symform system [Symform] comes with an alternative way of implementing distributed storage. It uses the advantages of a peer-to-peer solution to decrease storage costs and to provide supplementary security measure based on dispersion in addition to a standard data encryption.

Symform uses the RAID-96TM patented [Tabbara…11a], [Tabbara…11b], [Tabbara…14] technology for data protection and availability. Before being stored in the Symform cloud, the data stored in one dedicated folder at user device is divided into 64MB blocks and encrypted using the AES-256 algorithm. The unique encryption key for each block is the hash of the block itself. Data is encrypted at folder level, so the technique allows a de-duplication of data inside the client folder without need for decryption: if a block already exists, it will not be up-loaded in the cloud one more time. After encryption, each block is shredded into 64 fragments of 1MB each. Then, 32 parity fragments are added to every block using the Reed-Solomon error correction code. This processing results in 96 fragments corresponding to one block of the original data. These fragments are randomly distributed across 96 randomly chosen devices. To reconstruct protected data, 64 out of 96 fragments need to be assembled.

The combination of encryption and fragmentation provides a strong computational security. In order to decrypt one single block of data, a malicious user has to find the location of the 64 pieces of information, collect them from the storing devices and then break the AES-256 encryption, for which the key is twice as long as the AES-128 recommended by the NIST. This seems to be an insurmountable effort.

The weak point of the Symform system is its Cloud Control element responsible for the management of fragment locations and encryption keys. Symform user must totally trust their cloud provider or encrypt his data by himself before sending it to the cloud. Another problem is the fact that such system configuration is not resistant to attacks coming from Symform's insiders, such as malicious employees.

### 7.3. Exploiting data structures, confidentiality and machine trustworthiness

In case of a known data structure, fragmentation process can be improved by taking into account the uneven need for secrecy of different types of data. This way, confidential data can be separated from non-sensitive information and the storage place of both parts can be appropriately adapted. There is no sense in providing special secure architecture for the purpose of storage of piece of data, which do not reveal anything confidential. This idea was adopted by the authors of the object-oriented Fragmentation-Redundancy-Scattering [Fabre…94], [Fray…86] technique at the end of the last century. Then it has been modified to suit the cloud technology and the database as a service concept [Hudic…12], [Bkakria…13], [Vimercati…13].

The need of user interaction during decomposition process is one of the biggest problems of fragmentation of structured data. Each set of data types will have its own rules to define what is confidential and what is not. Moreover, it is possible that the combination of two non-confidential data fragments will reveal information about the confidential one. This was especially pointed out in [Ciriani…10].

#### 7.3.1. *Fragmentation-Redundancy-Scattering*

Fragmentation-Redundancy-Scattering (FRS) [Fabre…94] pro- vides accidental and intentional fault tolerance. In a first step, confidential objects belonging to an application are fragmented using a redundant algorithm until being bro- ken into pieces which do not reveal anything sensitive. Redundancy is achieved by the use of error processing techniques (like error correcting codes) or by anticipating the application design at the early stage of designing objects. At the end, the fragmented data is scattered over various workstations. Leftover fragments, which are still holding confidential information after the first step, are encrypted or stored on trusted devices. All remaining pieces are distributed over untrusted sites. Data processing, fragmentation, and defragmentation can be performed only on trusted sites. After almost two decades, the FRS technique from [Fabre…94] has been implemented in [Chougoule…11] using .net remoting.

### 7.3.2. Database fragmentation

Database as a service (DBaaS) [Cloud_Database] delivers similar functionality to classic, relational, or NoSQL database management systems in parallel to providing flexibility and scalability of an on-demand platform. The DBaaS user does not worry about the database provisioning issues, as the cloud provider responsibility is to maintain, back-up, upgrade and handle physical failures of the database system. Therefore, simplicity and cost effectiveness are the biggest advantages of such solution.

However, data owner loses control over outsourced data. This creates new security and privacy risks, especially when the concerned databases contain sensitive data, such as health records or financial information. In consequence, securing database services has become a need of paramount importance. A straightforward solution to the problem involves the encryption of the data before sending it to the storage provider. Large overhead and query processing limitation are the main drawbacks of such an approach. K-anonymization [Sweeney 02], t-closeness [Li 07] and l-diversity [Machanavajjhala 07] anonymization techniques are seen as a way of minimizing these inconveniences. In France, works on this subject has been recently done mainly for health data [Bergeat…14].

Database fragmentation promises an interesting alternative to database full encryption or full anonymization. One of the first works on the subject [Aggarwal…05] introduces a distributed architecture for preserving data privacy. A trusted client communicates with the end-user and utilizes two untrusted servers belonging to different storage providers ensuring physical separation of the information to be protected. By construction, storage providers do have access to the information that users entrust them with. Even if they are well aware that they should not incorrectly interact with the user's data and its integrity without endangering their own business, it is a common assumption to suppose them to be "*honest but curious*": they have the ability to observe, move, and replicate stored data, especially behind the virtualization mechanism. In [Aggarwal…05], outsourced data is partitioned among the two untrusted servers in a way that content at any one server do not breach data privacy. In order to obtain valuable information an adversary must gain access to both databases. Similarly, the system is also protected from insider attacks and the curiosity of the providers as long they do not ally together. Moreover, queries involving only one of the fragments are executed much more efficiently than on encrypted data. Nevertheless, this privacy-preserving outsourcing solution has a serious limitation. It assumes that the two servers are not allowed to communicate with each other. In a real world scenario such a condition can be hard to guarantee.

Another work [Ciriani…10], [Vimercati…13] comes with a solution for protection of confidentiality of sensitive information, based on mixing encryption and fragmentation. It defines **confidentiality constraint** as a subset containing one or more of relation attributes. Constraint involving only one attribute implies that value of the constraint attribute is sensitive and the only way of protecting it is the use of encryption. On the other hand, multi-attributes constraints specify that only association between the attributes of a given constraint are sensitive. In that case, there is no need to encrypt all the attributes values, because confidentiality can be ensured by fragmentation.

In [Ciriani…10], three scenarios of fragmenting a relation are presented. In the first one, a relation is divided into two fragments, which does not contain sensitive combination of non-encrypted attributes. The second scenario splits the relation into multiple fragments in a way that a query can always be evaluated on one of the fragments: each fragment contains non-encrypted attributes that does not violate the confidentiality constraints, as well as the encryption of all other attributes. Last fragmentation scenario avoids the use of encryption by introducing a trusted party (belonging to the data owner) for the storage of sensitive portion of data.

For each scenario, the authors associate fragmentation metrics supporting the definition of an appropriate fragmentation algorithm. Fragmentation metrics can aim at minimizing the number of fragments, maximization of affinity between attributes stored in one fragment or minimization of querying costs.

Bkakria's work [Bkakria…13] generalizes the approach presented in previous paragraphs to a database containing multiple relations. It introduces a new confidentiality constraint for protection of relationship between tables. Sensitive associations between relations are secured by the protection of primary key/foreign key relationships and separation (called vertical fragmentation) of the involved relations. Relations are transformed into secure fragments in which subsets of attributes respecting confidentiality constraints are stored in plaintext, while all others are encrypted. It introduces a parameter for evaluating the query execution cost and proposes a query transformation and optimization model for executing queries on distributed fragments. It also concentrates on the issue of preserving data unlinkability while executing queries on multiple fragments. Indeed, providers can collude with each other and then deduct information by observing query execution. To avoid such situation, [Bkakria…13] proposes the use of an improved Private Information Retrieval [Olumofin…10] technique, which allows querying a database without revealing query results to service providers. Results of implementation of the proposed approach are presented. Although the modified PIR solution is much faster than its predecessor, the processing time of record retrieval from multiple fragments is considerably slower in comparison to querying a single fragment.

The idea of splitting database into fragments stored at different cloud providers was also proposed in [Hudic…12]. In this approach, a database is first normalized and then different security levels (high, medium, low) are attributed to relations. Basing on these three levels and specific user requirements, data is encrypted, stored at local domain or distributed between providers.

All of the fragmentation methods presented in this subsection remain limited in terms of number of fragments. Moreover, proposed fragmentation algorithms in each case require user interaction in order to define data confidentiality level.

### 7.4. Conclusion

In previous sections we analyzed existing distributed storage systems providing additional secrecy by use of fragmentation. We also presented database fragmentation solution separating data in order to avoid full encryption. Few systems focus on providing a long term, secure and non-costly storage of data. Another motivation is the possibility of minimizing encryption inside databases, while still providing a good level of data protection.

### 7.4.1. Some issues and recommendations

In order to design an efficient storage system for fragmented data some problems still need to be overcome.

First, a process of fragment dispersion requires the assurance of securing data separation. A situation where data is fragmented, but where we would not or could not control where fragments are stored has to be considered as a *weak* situation. Using multiple independent providers can be a rapid and coarse-grained solution ([Aggarwal…09], [Hudic…12]), however, it can entail significant latency costs. An alternative could be the storage of data at one single provider site with the guarantee of physical separation. Unfortunately, the majority of cloud providers use virtualization, which prevents an end-user from such a control. With the development of bare-metal [Bare-metal] clouds like Trans-Lattice Storm [TransLattice_Storm], Internap [Internap] or Rackspace OnMetal [Rackspace], we believe it could be possible to control physical location of outsourced data within a single provider.

Second, there is a lack of published results showing performance allowing comparison between dispersed storage systems and the most common ones. In any case, fragmenting for protection always increases latency inside a system. Nevertheless, a good fragmentation technique must be combined with parallelization of processing to reach acceptable overall performance. It also should take into consideration the uneven levels of confidentiality requiring different level of protection for a collection of fragments. We have seen in this report (published in [Qiu…15]) how these ideas have been successfully utilized and developed for selective encryption of images using a General Purpose GPU (GPGPU).

Long term storage have a double issue with key management: the keys can be lost, the keys can no longer protect the data due to the constant progresses of hardware efficiency and cryptanalysis.

Last, but not least, fragmenting structured data strongly benefits from user interaction for definition of the confidentiality levels and depends a lot on the nature of the dataset. Designing an algorithm for separating confidential data from non-sensitive pieces would make the storage process much faster and easier to use.

## 8. Future Work

The results presented in this report are "work in progress" and we are just getting started. We have seen that due to constant progress in hardware architectures and cryptology these results are and will be constantly evolving. We defined a number of high level requirements at the beginning of this report. Some have been met. Moreover, we feel that more work remain to be done and is foreseeable to mature our results to satisfy a wider range of applications.

### 8.1. More security, more security analysis

We analyzed the level of security of the different FED methods with various statistics be correlation or entropy functions. We feel that it still remains to be able to define some formal method in order to compare these various methods and set their parameters in an optimal fashion with regard to the desired security level. It also remains to consider, test, and benchmark specific attack resistance, and understand possible strength and vulnerabilities of these methods. For instance, we have to better understand the effect of dispersion or the usage of decoys on security, none of the security analysis takes this aspect of our method into account. We also have to provide with at least a methodology and recommendations on protecting the map associated with a given fragmentation.

### 8.2. Agnosticism

We claimed agnosticism for our methods. However, we still need to run a number of experiments associated with security analysis for diverse data formats and nature such as high definition image resolutions like HD, truecolor or widely used compressed image standards like JPEG, NPG. This should be a simple extension of the work presented in section 6. Addressing audio and video formats is key in order to propose a more general system to protect all kinds of information. We also have to study various use cases, for instance, it is worth noticing that our method can also be used for soft encryption purposes.

### 8.3. Memory, performance, energy, and benchmark

Memory occupation and performance are critical and greatly influence the choice of one method versus another. We have seen the influence of the hardware architecture over the software architecture, particularly when using GPU with their issue of performance portability. We defined a benchmark from an end-user viewpoint for this very reason. Today, we need to be vigilant about technology progress since best in class solution is platform dependent. Using multiple processors is calling for considering energy consumption, and we are convinced that some energy can be easily saved without degrading performance.

### 8.4. Fragmentation and transmission

We have been studying fragmentation for protecting data during its storage. It remains to look at using this technique for protecting data during its transmission along all kind of networks. Early, analyses of this question are quite encouraging.

### 8.5. Defragmentation avoidance

We have seen that FHE had poor performance that fundamentally hinders its deployment. We are interested in looking once more in multiparty computation or linear secret sharing (LSS) ([Archer…2015]) which looks so promising despite the same hurdle.

### 8.6. Towards a general architecture

We designed and implemented several methods combining security with resilience while optimizing storage space and costs. They are agnostic supporting all kind of data of any nature or structure. Then, we will concentrate on overcoming performance issues by the use of parallel processing. As a final and most important step, we see the data separation along levels of confidentiality in order to address scalability.

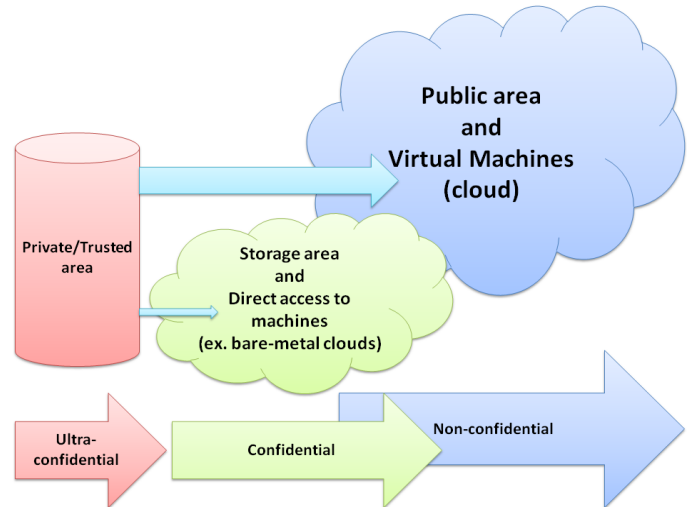

**Figure 28: Data storage principle regarding multilevel data confidentiality**

Figure 28 is a simplified version of Figure 1: 'Cone of confidentiality': Data confidentiality, machine trustworthiness, and data size. It shows the principle driving the architecture of our system. It can be considered as the projection of the 'cone' described in Figure 1 onto the (trustworthiness, confidentiality) dimensions. In a first approach, we divide the data into three categories: ultra-confidential, confidential and non-confidential (some banking organizations have four levels). Storage cost effectiveness is addressed by adapting the type of storage architecture to the level of confidentiality of these data. A trusted area, which will usually belong to a data owner who is in control of every process and data hosted in this area, will be used to store the most sensitive data as well as area of defragmentation. The rest of the data may go to two types of clouds. The non-confidential part of information will be kept inside popular public clouds that use virtualization with a cost effective contract. Information that can potentially reveal a secret may go to bare metals clouds, which provides more control of data location and isolation than the virtualization-based approach. Obviously, we are looking at minimizing the amount of data to be stored inside the most expensive zone and minimizing overall storage cost.

Finally, our ultimate goal is to design an architecture which is at the same time cost effective, efficient in terms of performance and effective in terms of protection. This solution will be intensively enabled by the existence of a large numbers of servers organized in public clouds and private distributed systems.

## 9. General Conclusion

The usual response when it comes to support data protection and privacy, is to transform (encryption using a key being a special case) the information to keep it unreadable or partially unreadable by someone who has not enough right and has not been given the way to inverse the said transformation. The usual response to guarantee some data resilience is to replicate it in order to keep information recoverable and available despite failures or attacks. After all, Mother Nature is using similar processes with encoding and replicating precious DNA molecules. The problem comes with scalability and duration: the need to both protect and keep safe large amount of information possibly for a long period of time. With massive amount of data or with all kind of continuous data streams production, costs of traditional protection by full encryption processes or of resilience by sheer data replication increase too rapidly. This is true either in term of performance and energy consumption relatively to encryption processes or in term of memory relatively to the replication processes. Our current work is showing that this problem can be addressed in a three prong approach: fragmentation and massive parallelization (in particular, exploiting GPUs), encryption of information along gradual level of confidentiality, and at last, dispersion of the said fragments over a large network of machines potentially organized into one or multiple clouds with various levels of trustworthiness.

Today, we feel that the concept fragmentation has started to mature and is seriously considered by the industry: IBM recently bought out Cleversafe one of the technologies we presented in Table 17 for $1.3B[4]. We feel that the concept of dispersion deserves more attention and more experimentation. It is quite clear that the way fragments are distributed over different physical devices is bringing more or less additional protection and robustness to the system; at the same time, it can bring more or less latency. It is quite clear that dispersing data over different providers, in particular between private and public areas can bring more or less cost effectiveness.

Our work started with the strong belief that the nature of the data ought to be taken into consideration and algorithms ought to be designed and specialized according to the fact that data is an image, a video, or a text. We are ending up this project by delivering three algorithms agnostic with regards to the nature or type of data by both considering data not as a stream of zeros and ones but by considering data as a two dimensional matrix and dealing with integrity of the diverse transformations (recommending the usage of DWT/LeGall over DCT that we were using at the beginning of this study, DFT, or FFT) and cyphers. We believe that this unexpected result joined to the usage of GPU for performance will allow broadening the scope of applications particularly of selective encryption.

It was difficult to find related work in the literature (mostly for terminology reason). It was not that easy to convince oneself of the validity of the concepts that have been presented. The potential issues with performance and latency was constantly pregnant in our study. Today, our work is hopefully clearly showing that fragmentation, encryption, and dispersion can be seen as a general and potent process to efficiently protect data. We believe it is still early to see them widely used. Although it is still an open question to mathematically compare and position one security method versus another one, we gathered a set of security analysis which allowed us checking our three algorithms for security. Our objective was, starting from the briefly presented state of the art in section 7, to address few of the challenges considered in section 2 and 4 by designing a system that would be more efficient than just encrypting the information to be protected. We are finishing this project producing three new algorithms of fragmentation/defragmentation, and demonstrated their competitiveness against full encryption in terms of performance, scalability, resilience, and protection.

---

4 : https://www.bloomberg.com/news/articles/2016-02-24/ibm-paid-1-3-billion-to-acquire-cleversafe-in-hybrid-cloud-push

## 10. Acknowledgements

This report owes a lot to the works of two PhDs students: Qiu Han and Katarzyna Kapusta who designed and developed the algorithms presented in this report. Moreover, many thanks go to Patrick Lambein who refactored and developed some fragmentation/defragmentation code in Java in order to prepare the code to become open source and to Hassan Noura who introduced us to some elements of security analysis.

## 12. Annex: technical environment and access to the code

We provide here, the list of GPU architectures and various hardware platforms that have been used for developing code and experimenting on data protection. Follows the list of software used to perform transformations or encryption, then, the urls to access to the code.

**Hardware:**
Laptop,
GPU: Nvidia Nvs 5200M
CPU:Intel I7-3630QM
CPU:Intel I7-4770HQ (Apple Macbook)

Desktop,
GPU: Nvidia Geforce GTX 780
CPU Intel I7-4770K

Mobile,
Apple Cortex A8 Chip (Apple iphone6)

Teralab Hardware platform (https://www.teralab-datascience.fr/fr/)

**Software:**
Windows 7, Visual studio 2012, CUDA 5.5/6.0, Eclipse Luna, Matlab
Xcode 7.0 SDK (apple develop software platform)
DCT: Nvidia CUDA SDK: https://developer.nvidia.com/cuda-code-samples
DWT: Nvidia CUDA SDK: https://developer.nvidia.com/cuda-code-samples (*modified from other open-source project by using the CUDA SDK*)
AES: Cryptoo++: www.cryptopp.com
SHA-512, SHA-256: http://hashcat.net/oclhashcat/

Teralab Software platform

**Data:**
Test images used in this paper are very well known and can be found in many publications. Their bitmap format can be found under the names of 'Barbara' and 'baboon' in various open test image databases and are accessible from the web.
Data from the ITEA2 CAP project, from La Poste, the French Post Office.

**Code:**
The C / C++ code on selective encryption is available at:
https://www.dropbox.com/sh/p4mloc2hlp946kv/AADXLcnvWljvxwbybKqXoa5wa?dl=0
The Java code on fragmentation/defragmentation is available at:
https://github.com/kasiakapusta/fragmentation with the username: CAPmember and the password: CAPItea2Fragm